\documentclass[preprint,3p,times,twocolumn]{elsarticle}

\usepackage{amssymb}
\usepackage{graphicx}
\usepackage{subfigure}
\usepackage[inline, shortlabels]{enumitem}
\usepackage{amsmath}
\usepackage{url}
\usepackage{algorithmic}

\usepackage{array}
\usepackage{booktabs}

\newcommand{\secref}[1]{Section~\ref{#1}}

\newcommand{\figref}[1]{Fig.~\ref{#1}}    
\newcommand{\figsref}[2]{Figs.~\ref{#1}--\ref{#2}}    
\newcommand{\Figref}[1]{Figure~\ref{#1}}  
\newcommand{\tabref}[1]{Table~\ref{#1}}  
\newcommand{\Tabref}[1]{Table~\ref{#1}}  
\newcommand{\equref}[1]{Eq.~(\ref{#1})}
\newcommand{\eqsref}[2]{Eq.~(\ref{#1})--(\ref{#2})}

\newcommand{\etc}{etc.\ }
\newcommand{\eg}{e.g.,\ }
\newcommand{\Eg}{E.g.,\ }

\newcommand{\ie}{i.e.,\ }
\newcommand{\vs}{vs.\ }

\newcommand{\centigrades}{{\textdegree{}C}}

\journal{Applied Energy}

\begin{document}

\begin{frontmatter}

\title{Physics Informed Neural Networks for Control Oriented Thermal Modeling of Buildings}



\author{Gargya Gokhale\corref{cor1}}
\ead{gargya.gokhale@ugent.be}
\cortext[cor1]{Corresponding Author}

\author{Bert Claessens}
\author{Chris Develder}

\begin{abstract}
Buildings constitute more than 40\% of total primary energy consumption worldwide and are bound to play an important role in the energy transition process.
To unlock their potential, we need sophisticated controllers that can understand the underlying non-linear thermal dynamics of buildings, consider user comfort constraints and produce optimal control actions.
A crucial challenge for developing such controllers is obtaining an accurate control-oriented model of a building.
To address this challenge, we present a novel, data-driven modeling approach using physics informed neural networks.
With this, we aim to combine the strengths of two prominent modeling frameworks: the interpretability of building physics models and the expressive power of neural networks.
Specifically, we use measured data and prior information about building parameters to realize a neural network model that is guided by building physics and can model the temporal evolution of room temperature, power consumption as well as the hidden state, \ie the temperature of building thermal mass.
The main research contributions of this work are:
\begin{enumerate*}[(1)]
\item \label{it:contrib:NNs} we propose two new variants of physics informed neural network architectures for the task of control-oriented thermal modeling of buildings,
\item \label{it:contrib:training-efficiency} we show that training these architectures is data-efficient, requiring less training data compared to conventional, non-physics informed neural networks, and 
\item \label{it:contrib:accuracy} we show that these architectures achieve more accurate predictions than conventional neural networks for longer prediction horizons (as needed for effective control).
\end{enumerate*}
We test the prediction performance of the proposed architectures using both simulated and real-word data to demonstrate~\ref{it:contrib:training-efficiency} and~\ref{it:contrib:accuracy} and argue that the proposed physics informed neural network architectures can be used for control-oriented modeling.
\end{abstract}



\end{frontmatter}


\section{Introduction}
According to the recent IPCC report on climate change, global temperature is expected to reach the 1.5°C threshold in the next decades~\cite{ipcc-2021}. In the fight against climate change, the energy and power sector is going through numerous changes such as phasing out of coal-based generation, the addition of renewable energy sources and decentralization of generation and storage units. Concurrently, there is a growing need for efficient and flexible energy consumption that can accommodate the energy generated by intermittent renewable energy sources such as wind and solar power~\cite{why-dr}. An important sector for providing this energy efficiency and flexibility is the building sector. As of 2016, buildings accounted for 40\% of total primary energy consumption worldwide and around 55\% of total electricity consumption in the EU~\cite{build-consump-sota2016}. With these numbers expected to rise over the years, efficient control of building energy consumption will play a crucial role in the energy transition process. 

Significant research has been carried out in the context of control algorithms for energy management in buildings, ranging from simple Rule-based Controllers to advanced controllers like Model Predictive Control (MPC) and Reinforcement Learning (RL)~\cite{gen-review-2019}. In MPC, a physical model of the system is used to anticipate the future behavior of the system and optimize its performance~\cite{mpc-basic}. This enables MPC-based controllers to be sample efficient and produce interpretable control decisions. However, the accuracy of MPC is closely related to the fidelity of the model, which is often difficult to obtain for real-world scenarios~\cite{mpc-challenges}. Contrary to this, data-driven controllers like RL, work directly with past interactions between the system, without the need for explicit physics knowledge. Although these RL-based controllers have shown promising results, they present a black-box solution that requires large amounts of training data. Additionally, in previous work such as~\cite{deep-rl-1}, RL controller was trained using a physics model-based simulator to ensure that training data obtained was sufficiently diverse and to avoid taking harmful exploration actions.

This makes obtaining accurate building models a crucial requirement for developing better control algorithms. A variety of modeling techniques have been studied previously and are broadly classified into physics models (white box, grey box) and data-driven models (black box)~\cite{model-sota}. The physics models involve solving a system of partial differential equations based on the underlying physical laws, commonly achieved using numeric solvers such as EnergyPlus, Modelica, as presented in, \eg \cite{EP-mpc,modelica}. The use of such models however has been limited in the control domain, primarily due to the high computational cost associated with solving the underlying system of partial differential equations~\cite{EP-mpc}. Alternatively, a lumped parameter model using resistive and capacitive networks is used for control-oriented modeling. With this framework, different thermal components in a building are modeled using an RC network and simplified to obtain a lower order model that is easier to solve. However, even with these approximations, the models obtained are highly specific and require significant modeling effort as demonstrated in~\cite{mpc-real-problems}.

Data-driven models circumvent these modeling challenges by relying completely on obtained data. Previously, techniques such as ARIMA, Genetic Algorithms, Neural Networks, etc., have been studied and have shown good modeling capabilities~\cite{model-sota, data-driven-narx}. Yet, as discussed in~\cite{model-sota}, these techniques have their own challenges in the form of huge training data requirement and lack of interpretability. 

To get the best of both these worlds, we propose to incorporate self-learning, physics guided models with model-based reinforcement learning algorithms to develop interpretable control agents in a data-driven manner. As a first building block, we propose to work with Physics Informed Neural Network architectures to learn physically relevant control-oriented models of real-world systems. This is achieved by explicitly providing information related to the underlying physics of the system to a deep neural network during the training procedure.

The main contributions of this paper can be summarized as:
\begin{enumerate}[(1)]
    \item We propose two new variants of physics informed neural network architectures  (\secref{sec:math_model}) for control-oriented modeling of thermal behavior of a building and validate their accuracy using simulated data (\secref{sec:results:validation}).
    \item Based on real-world cold storage data (\secref{sec:exp-setup}) we show that these physics informed neural networks perform better than conventional, non-physics informed neural networks at predictions for longer time horizons (\secref{sec:results:horizon}).
    \item We further show that training these physics informed neural network architectures is a data-efficient process, requiring less training data than conventional, non-physics informed neural networks (\secref{sec:results:training-data-size}).
\end{enumerate}

While we acknowledge that the general concept of using physics informed neural networks models in itself is not new (as indicated through the literature review in the subsequent \secref{sec:literature}), the specific physics informed neural network model we have designed (incorporating basic constraints based on a simple RC model) is. Through aforementioned experiments we demonstrate our model's feasibility and practical applicability, based on experiments using real-world data. 


\section{Related Work}
\label{sec:literature}
This section presents a non-exhaustive review of previous work related to our paper. We specifically focus on
\begin{enumerate*}[(i)]
\item Building Control and Modeling Algorithms and
\item Physics Informed Neural Networks.
\end{enumerate*}

\subsection{Building Control and Modeling}
Extensive research has been carried out previously in this domain, with work such as \cite{gen-review-2019,mpc-review-2021}~presenting exhaustive reviews of different control algorithms. MPC has emerged as an established control technique with works such as~\cite{expt-mpc, building-model2012, swiss-mpc} presenting case studies for practical implementations in real-world buildings. These works show that MPC-based control strategies can lead to cost savings of about $20\%$ compared to the existing rule-based control algorithms. An MPC strategy involves a physical model of the system and a set of constraints to formulate a receding horizon optimization problem that is solved at every time step to obtain optimum control actions~\cite{mpc-basic}. Authors in~\cite{building-model2012, swiss-mpc}~utilize a grey-box RC model for the buildings. This leads to a bilinear building model and results in a non-linear optimization problem that can still be solved with reasonable accuracy using a sequence of linear programs~\cite{swiss-mpc}. Solving this optimization problem is computationally expensive and can limit the practical applicability of MPC controllers. Besides this, MPC controllers are expensive to obtain as indicated in~\cite{swiss-mpc}, whose authors conducted a cost-benefit analysis of using MPC-based control strategies in real-world buildings. They concluded that, while MPCs can lead to a decrease in operating costs, the investment costs are much higher, primarily due to higher costs associated with the modeling of buildings, thus prohibiting widespread commercial application. Further, these building models are seldom scalable and need to be developed for individual buildings. \Eg in \cite{mpc-real-problems}, authors discuss the model identification process for a real-world building and present a procedure for estimating building parameters. This procedure leads to models with accurate multi-step temperature predictions (0.3\centigrades~prediction error). However, this identification process involves solving a quadratic program to obtain good initial estimates of building parameters, followed by solving a multi-step prediction optimization problem to obtain the final model. This highlights the high computational requirements for obtaining a good building model and the lack of scalability. Additionally, the modeling approaches presented above approximate the non-linear thermal dynamics of the building using a first-order Euler discretization. As presented in~\cite{dual_estimate}, such explicit approximations can lead to inaccurate buildings models especially in scenarios involving low data sampling rates. This leads to approximate models which can be susceptible to errors and lead to biased control actions, as discussed in \cite{EP-mpc, mohak2020}.

Data-driven control techniques circumvent aforementioned shortcomings of MPC by completely relying on collected data as presented in~\cite{data-driven-predictive2021}. Owing to the recent success of works such as~\cite{alpha-go}, Reinforcement learning-based controllers are gaining importance in developing controllers for buildings~\cite{rl-review}. RL-based controllers are self-learning controllers that use data collected from past interactions between the system and the controller to learn the dynamics of the system and achieve a pre-defined objective~\cite{sutton-barto}. Works such as~\cite{deep-rl-1, rl-fqi-ql} have studied RL-based controllers in the context of building control and show that such RL controllers can lead to $5-12\%$ energy savings compared to the existing rule-based controllers. Additionally, \cite{mpc-vs-rl} compares the performance of MPC and RL controllers to show that RL controllers are able to outperform a linear MPC-based controller for two different test scenarios. Though these works indicate promising results for RL-based controllers, they also highlight existing challenges in real-world deployment of RL. These include large training data requirement, lack of interpretability, need for safe explorations, etc.~\cite{rl-challenges}. \Eg in \cite{rl-fqi-ql}, the authors use one year of data for training the RL controller using random explorations. Similarly, in \cite{deep-rl-1}, the authors use 2 months of temperature data for obtaining a training data size equivalent to $3000$ simulated trajectories. This data intensive nature and need for significant exploration represents a common challenge faced by RL-based control strategies. Hybrid control approaches have been studied to mitigate some of these problems by combining domain knowledge with these RL controllers~\cite{mohak2020, trajectory-bert}. \Eg in \cite{trajectory-bert}, the authors present a hybrid control strategy by merging model-free and model-based control strategies. They propose an aggregate-and-dispatch control framework for a cluster of water heaters in which an MPC controller calculates energy set-points for the cluster and the dispatch is carried out based on a fitted Q-iteration RL strategy.

We present a different approach, where instead of using a model of the system directly, we focus on learning this model using the available data and then using it in a model-based RL approach. While different techniques for this control-oriented modeling problem have been studied previously, these techniques were focused on creating convex, linear (or bi-linear), time invariant models compatible with MPC formulation and available optimization solvers~\cite{control-models, control-models-2}. In contrast, our objective in this work is to learn a low dimensional, latent space dynamics model of the system to use it in RL, where these latent representations can be used to learn optimum control policies as demonstrated in~\cite{mu-zero2020, dreamer}. Concretely, we propose using Physics Informed Neural Network architectures~\cite{pinn-og}.

\subsection{Physics Informed Neural Network Architectures}
As introduced in \cite{pinn-og}, Physics informed neural networks represent a novel class of neural network architectures where prior knowledge about the system is encoded explicitly in the architecture. This work is similar to \cite{hamiltonian}, where inductive biases based on the underlying physics laws are coded directly into the network. Several works have built upon this idea and have shown promising results in obtaining approximate solutions for difficult physics problems such as two body mechanics~\cite{pinn-idlab-2021}, heat transfer~\cite{pinn-heat-transfer2021}. In \cite{pinn-og, pinn-heat-transfer2021}, the encoded physics knowledge is strictly enforced on the predictions of the neural network and assumes the availability of complete physics. Differing slightly from this approach, in \cite{pinn-idlab-2021}, the authors enforce partially known physics and learn remaining physics parameters using the available data. These approaches show that trained models are better at extrapolating and require fewer training samples.  Consistent with these works, we propose using physics informed neural networks for modeling the thermal behavior of a building. This is an emerging domain in building energy modeling and control, with previous works such as~\cite{physARMAX, phys_consistent, pinn-multi-zone} presenting different approaches for providing prior physics knowledge to conventional black-box algorithms. In~\cite{physARMAX}, the authors present a Physics-informed ARMAX method for modeling buildings and using these models with MPCs. The results indicate good performance of MPCs based on this model, including significant training sample efficiency attributed to the prior knowledge that is provided to the physics-informed ARMAX models. In~\cite{phys_consistent}, the authors present a physically consistent neural network architecture comprising of a physics-informed module in parallel to a black box module. With this approach, the authors show good prediction performance for longer prediction horizons as compared to a grey-box model. In~\cite{pinn-multi-zone}, the authors present a physics-constrained deep learning model for control oriented model of a commercial building. The authors propose a structured recurrent neural dynamics model that models the non-linear thermal dynamics of the building using individual linear neural blocks with constrained eigenvalues. The authors show that such an architecture is data efficient, requires less training time and gives state-of-the-art prediction results for the case of large office buildings. \\
Differing from the works presented above, we encode the physics directly by using neural network outputs to calculate additional physics-based losses. Additionally, we propose to extract low-dimensional latent representations which correspond to the hidden states of the system and use prior physics knowledge to guide them towards a physically relevant space. This ensures that the obtained latent representations are disentangled, as opposed to the unsupervised learning cases discussed in~\cite{disentangle1, vae-disentangle}. Once trained, these physics informed neural network models can be used with model-based RL algorithms such as MuZero~\cite{mu-zero2020}, Dreamer~\cite{dreamer} to obtain optimum control actions for building control.
\\


\section{Mathematical Modeling}
\label{sec:math_model}

With a focus on obtaining control-oriented models, we first model the thermal behavior of the household as a discrete-time Markov Decision Process (MDP), a commonly used sequential modeling framework~\cite{sutton-barto}. Subsequently, we train our physics- informed neural networks to predict the `next state’ given the `current state’ and `action’. This modeling approach has been discussed in this section along with the formulation of Physics informed neural networks. 

\subsection{Problem Formulation}
\label{sec:mat_model:problem}

A  Markov Decision Process (MDP) is a commonly used framework to model sequential decision making problems~\cite{sutton-barto}. An MDP consists of 4 main components: state space ($\mathbf{X}$), action space ($\mathbf{U}$), state transition function ($f$) and reward function ($\rho$). In a fully observable setting, the transition function $f\colon \mathbf{X}\times \mathbf{U}\times \mathbf{W}\to\mathbf{X}$ represents the true mapping at time step $i$, between the current state ($\textbf{x}_i$), the current action ($\textbf{u}_i$), an exogenous parameter ($\textbf{w}_i$) and the next state ($\textbf{x}_{i+1}$) of the system and is given as:
\begin{equation}
    \begin{split}
        \textbf{x}_{i+1} = f(\textbf{x}_i,\textbf{u}_i, \textbf{w}_i),\\
    \end{split}
\label{eq:transition_f}
\end{equation}
Here, $\textbf{w}_i$ represents the stochasticity in the system and is assumed as an independent random variable. The transition function ($f$) represents the true dynamics of the system and to obtain a control-oriented model, it is necessary to approximate this transition function. For data-driven methods, this reduces the problem into a supervised learning problem with the objective of estimating the transition function using a labeled set of state transitions $\mathcal{F}=\{(\textbf{x}_1,\textbf{u}_1, \textbf{w}_1,\textbf{x}_2), \ldots, (\textbf{x}_N,\textbf{u}_N,\textbf{w}_N, \textbf{x}_{N+1}) \}$. \Eg a neural network with parameters $\theta$ can be trained using Stochastic Gradient Descent~(SGD) to solve the following optimization problem:
\begin{equation}
    \min_{\theta} \frac{1}{N} \sum_{i=1}^{N} ({\textbf{x}}_{i+1} - f_{\theta}(\textbf{x}_i,\textbf{u}_i,\textbf{w}_i))^{2}.
\label{eq:NN-Loss}
\end{equation}
With this conventional approach, the neural network learns to estimate the transition function by fully relying on the training data-set, without explicitly learning the underlying physical relationships. It should be noted that \equref{eq:NN-Loss}~represents the scenario for a fully observable system, where complete state information is known and can be used to obtain predictions for the next states. However, a real-world system; such as a thermal model for a building; is generally partially observable where some state parameters cannot be measured or obtained directly. In such cases, using \equref{eq:NN-Loss} directly is not useful as the observed states lack complete information. To mitigate this, \cite{sutton-barto} presents different approaches, one of which involves engineering new high-dimensional features based on the observed states. \Eg in case of a thermal model for a building, the observed state can include measurements of room temperature or actual power consumption, whereas a hidden state parameter can be the temperature of building thermal mass (\eg~walls, furniture~\etc) which is difficult to measure or estimate accurately. Accordingly, to compensate for this missing state parameter, a sequence of past room temperature measurements can be used in lieu of a single room temperature measurement and this corresponds to an engineered feature for mitigating partial observability of this system.

For such partially observable MDPs, the state space($\textbf{X}$) consists of an observable component ($\textbf{X}^{\text{obs}}$) and a feature engineered component ($\textbf{X}^{\text{f}}$) such that $\textbf{X}= \textbf{X}^{\text{obs}} \times \textbf{X}^{\text{f}}$. With this high dimensional state representation, a neural network can be trained to estimate the next observable state ($\textbf{x}^{\text{obs}}_{k+1}$) given state, action and other exogenous inputs, thus modifying the optimization problem in \equref{eq:NN-Loss} as:
\begin{equation}
\begin{split}
    \hat{\textbf{x}}^{\text{obs}}_{i+1} &= f_{\theta}(\textbf{x}_i,\textbf{u}_i, \textbf{w}_i),\\
    \min_{\theta}\frac{1}{N} \sum_{i=1}^{N} &({\textbf{x}}^{\text{obs}}_{i+1} - f_{\theta}(\textbf{x}_i,\textbf{u}_i, \textbf{w}_i))^{2}.
\end{split}
\label{eq:pomdp-loss}
\end{equation}
Aside from this data-driven approach, for problems where the physics of the system are known apriori, the system dynamics can be approximated using a system of ordinary or partial differential differential equations that relate the observable states ($\textbf{x}^{\text{obs}}_{i}$), actions ($\textbf{u}_i$), other exogenous factors ($\textbf{w}_i$), hidden state parameters ($\textbf{z}_{i}$) and system parameters ($\Omega$). This is represented using a generic differential operator ($\mathcal{D}_{\Omega}$) as:
\begin{equation}
    \mathcal{D}_{\Omega} (\textbf{x}_i, \textbf{u}_i, \textbf{z}_{i}, \textbf{z}_{i+1},\textbf{w}_{i}) = 0.
\label{eq:differential_operator}
\end{equation}
In the following section, we present the physics informed neural network architectures, which combine \eqsref{eq:pomdp-loss}{eq:differential_operator} to obtain control-oriented models for systems where the underlying physics are known a priori. 

\subsection{Physics Informed Neural Network Architectures}
\label{sec:math_model:PhyNN}

Consistent with previous works such as~\cite{pinn-og, pinn-idlab-2021}, we explicitly encode the underlying physics of the system in a standard neural network architecture and then train it on the data collected. Assuming a partially observable setting, the network is trained to predict the next observable state~($\textbf{x}^{\text{obs}}_{i+1}$) and a latent representation~($\textbf{z}_{i}$) using a high dimensional state input~($\textbf{x}_{i}$), action~($\textbf{u}_{i}$) and exogenous information~($\textbf{w}_{i}$). This is done by setting up a constrained optimization problem based on \equref{eq:pomdp-loss} as:
\begin{equation}
\begin{split}
    &\min_{\theta, \Omega} \ \frac{1}{N} \sum_{i=1}^{N} \left( \, \textbf{x}_{i+1}^{\text{obs}} - \hat{\textbf{x}}_{i+1}^{\text{obs}} \, \right)^{2}  \\
    &\text{s.t.} \ \ \mathcal{D}_{\Omega} (\textbf{x}_i, \textbf{u}_i, \hat{\textbf{z}}_{i}, \hat{\textbf{z}}_{i+1}, \textbf{w}_{i}) = 0, \\
    & \forall \ \ (\textbf{x}_i, \textbf{u}_i, \textbf{w}_{i}, \textbf{x}^{\text{obs}}_{i+1}) \in \mathcal{F}.
\end{split}
\label{eq:PINN-opt}
\end{equation}
To solve this optimization problem, we define a loss function composed of two terms; $\mathcal{L}_{\text{reg}}$ represents the mean squared error loss for regression and $\mathcal{L}_{\text{phys}}$ represents the physics-based loss that makes the network adhere to the underlying physics. It should be noted that for real-world scenarios the dynamics of the system is influenced by exogenous parameters/ noise and does not exactly satisfy~\equref{eq:differential_operator}. To manage this, we replace the strict equality by a least-squared error term as shown in the loss term formulation in~\equref{eq:gen-loss}.
\begin{equation}
\begin{split}
    \text{Loss} &= \mathcal{L}_{\text{reg}} + \lambda \> \mathcal{L}_{\text{phys}} \\
    \mathcal{L}_{\text{reg}} &= \frac{1}{N} \sum_{i=1}^{N} \left(\, \textbf{x}_{i+1}^{\text{obs}} - \hat{\textbf{x}}_{i+1}^{\text{obs}}\,  \right)^{2}\\
    \mathcal{L}_{\text{phys}} &= \frac{1}{N} \sum_{i=1}^{N} \left( \, \mathcal{D}_{\Omega}^{\phantom{b}} (\, \textbf{x}_i, \textbf{u}_i, \hat{\textbf{z}}_{i}, \hat{\textbf{z}}_{i+1}, \textbf{w}_{i}) \, \right)^{2} \\
\end{split}
\label{eq:gen-loss}
\end{equation}
The influence of the physics-based loss term is regulated using $\lambda$. The optimization problem defined in \equref{eq:PINN-opt} can then be solved by using stochastic gradient descent to minimize the loss defined in \equref{eq:gen-loss}.

\begin{figure}[ht]
\centering
\subfigure[PhysNet]{\includegraphics[width=3.0in]{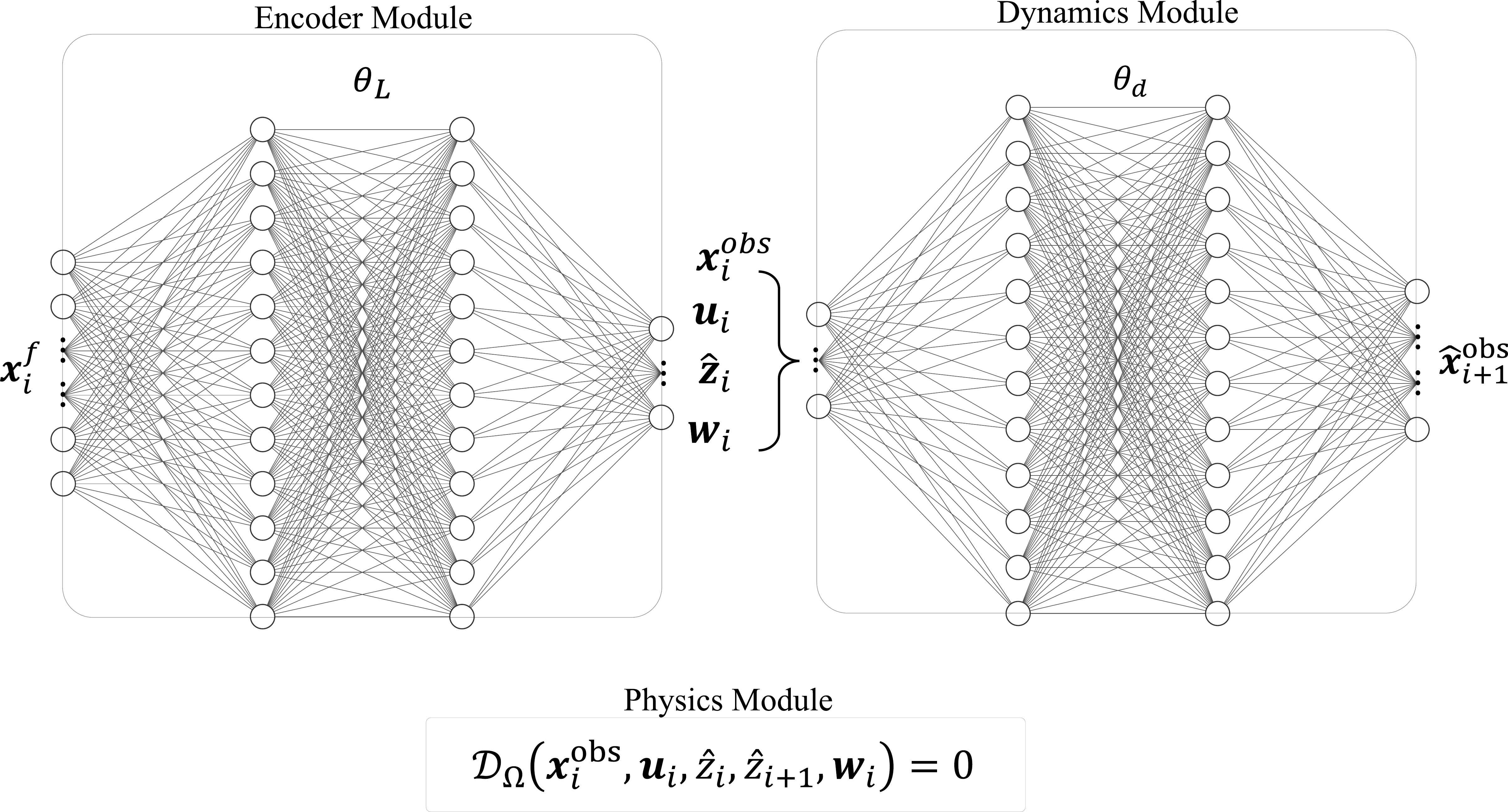}%
\label{subfig:encoder-pinn}}
\hfil
\subfigure[PhysReg MLP]{\includegraphics[width=2.1in]{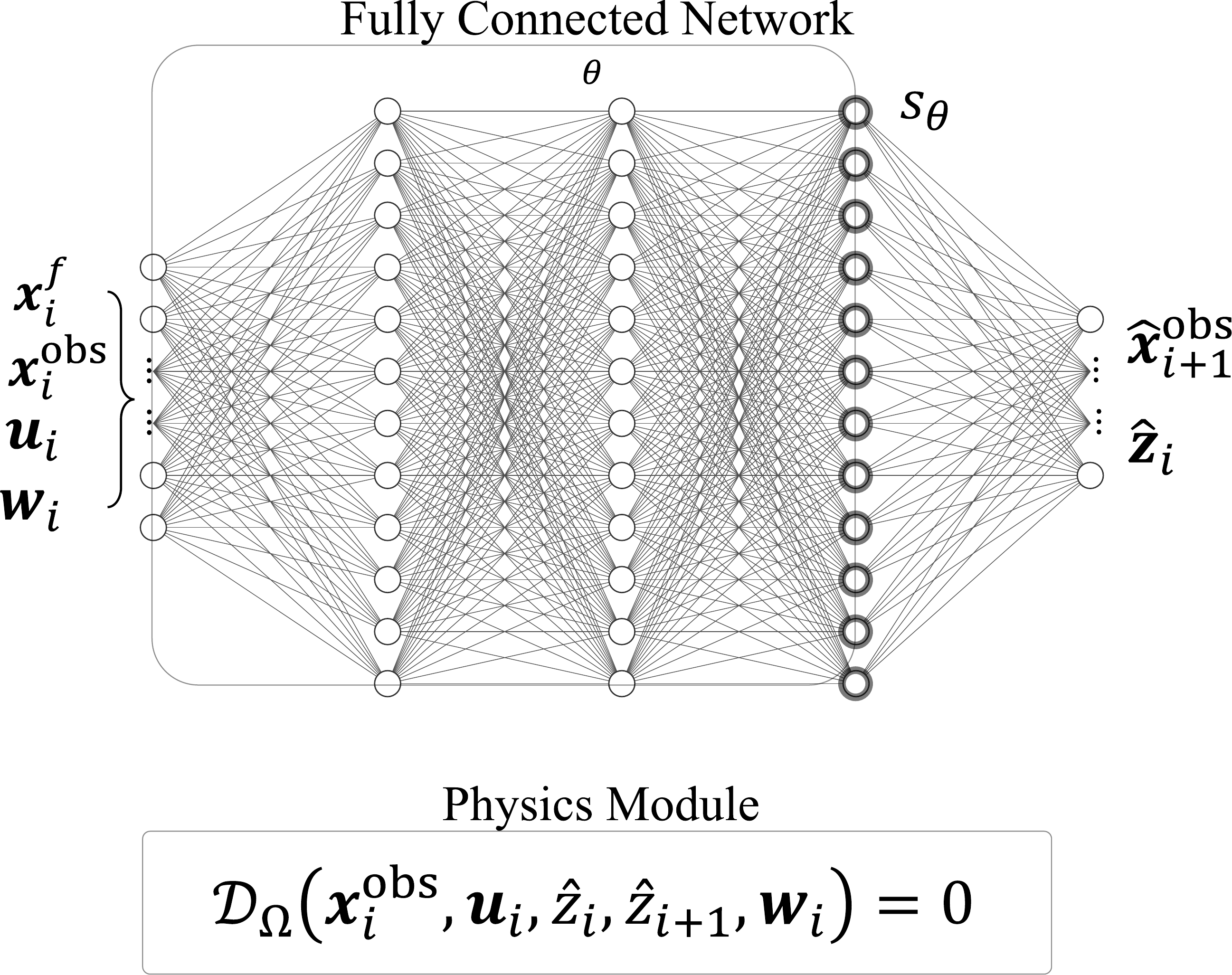}%
\label{subfig:vanilla-pinn}}
\caption{Physics Informed Neural Network architectures. $\textbf{x}^{\text{f}}_i$ represents a high dimensional feature engineered state component of the system, whereas $\textbf{x}^{\text{obs}}_{i}$ represents the observable components of the input sample $i$. $\textbf{z}_i$ represents a low-dimensional latent representation of the system.}
\label{fig:pinn-arch}
\end{figure}

Based on this formulation, we propose two variants of Physics Informed Neural Networks as shown in \figref{fig:pinn-arch}. The proposed networks have different architectures but are both trained based on the methodology described in \eqsref{eq:PINN-opt}{eq:gen-loss}. 
\Figref{subfig:encoder-pinn} presents an architecture comprising two modules, an Encoder and a Dynamics module. The encoder module is parameterized by $\theta_{\text{L}}$ and the dynamics module is parameterized by $\theta_{\text{d}}$. The encoder module creates a bottleneck and encodes the high dimensional, feature engineered component of state inputs~($\textbf{x}^{\text{f}}_i$) into a low dimensional latent representation~($\hat{\textbf{z}}_{i}$). This latent representation along with observable state information~($\textbf{x}_{i}^{\text{obs}}$), action~($u_i$) and other exogenous information~($\textbf{w}_{i}$) are then used by the dynamics module of the network to predict the next observable state~($\hat{\textbf{x}}_{i+1}^{\text{obs}}$) of the system. Thus, a forward pass of this network can be expressed as:
\begin{equation}
\begin{split}
    \hat{\textbf{z}}_{i} &= g_{\theta_{L}}(\textbf{x}^{\text{f}}_i),\\
        \hat{\textbf{x}}_{i+1}^{\text{obs}} &= h_{\theta_{d}}(\hat{\textbf{z}}_{i}, \textbf{x}_{i}^{\text{obs}}, u_{i}, \textbf{w}_{i})
\end{split}
\label{eq:physnet-forward}
\end{equation}
With this architecture, the prediction for next observable state depends on, among other parameters, the prediction of the latent representation~($\hat{\textbf{z}}_{i}$). This ensures that the encoded representation obtained from this network contains information regarding the dynamics of the system. 

Differing slightly from this approach, \figref{subfig:vanilla-pinn} presents a conventional fully connected neural network architecture where physics knowledge is incorporated based on~\eqsref{eq:PINN-opt}{eq:gen-loss}. The inputs of this architecture comprise of the full state representation~($\textbf{x}_{i} = (\textbf{x}^{\text{f}}_i, \textbf{x}_{i}^{\text{obs}})$), action and exogenous information. With these inputs, the network predicts the next observable state and a latent representation simultaneously. Thus, there is explicit parameter sharing between these predictions and these shared parameters are represented by $\theta$. The output layer uses an identity activation function and hence the outputs ($\hat{\textbf{x}}^{\text{obs}}_{i+1}$ and $\hat{\textbf{z}}_{i}$) are linear combinations of output of the last hidden layer of the network ($s_{\theta}$). Thus, a forward pass of this network can be formulated as:
\begin{equation}
\begin{split}
    \hat{\textbf{z}}_{i} &=  \textbf{g}_{1} s_{\theta}(\textbf{x}_i, \textbf{u}_{i}, \textbf{w}_{i}) + \textbf{g}_{2},\\
        \hat{\textbf{x}}^{\text{obs}}_{i+1} &= \textbf{h}_{1} s_{\theta}(\textbf{x}_i, \textbf{u}_{i}, \textbf{w}_{i}) + \textbf{h}_{2},
\end{split}
\label{eq:physreg-forward}
\end{equation}
where $\textbf{g}_{1}$, $\textbf{g}_{2}$, $\textbf{h}_{1}$ and $\textbf{h}_{2}$ are matrices of appropriate dimensions. With this parameter sharing, the obtained latent representation does not contribute in obtaining the predictions for the next observable state. Consequently, this latent representation may not contain sufficient information about the dynamics of the system. However, due to the physics, the latent representation is physically relevant and represents the hidden parameters of the system. Hence, in this case, the physics module acts as a regularization term guiding the network to learn the dynamics of the system and some latent representations simultaneously.


\section{Experimental Setup}
\label{sec:exp-setup}
In this section, we apply the general methodology introduced in \secref{sec:math_model} for the case of thermal modeling of a building. We will detail the thermal building model used, the type of experiments performed and the configurations of physics informed neural network architectures used. We compare the prediction accuracy of both architectures against a similar conventional neural network and assess whether the proposed architectures can be used for control applications.

\subsection{Thermal Model of a Building}

A simplified scenario is considered with a single room (or zone) that is heated using a heat source.
To model this scenario, we adopt a grey box modeling approach using the 2R2C network model~\cite{Vrettos2018}, illustrated in \figref{fig:rcnet} and with the following state-space formulation: 
\begin{figure}[t]
    \centering
    \includegraphics[width=.7\linewidth]{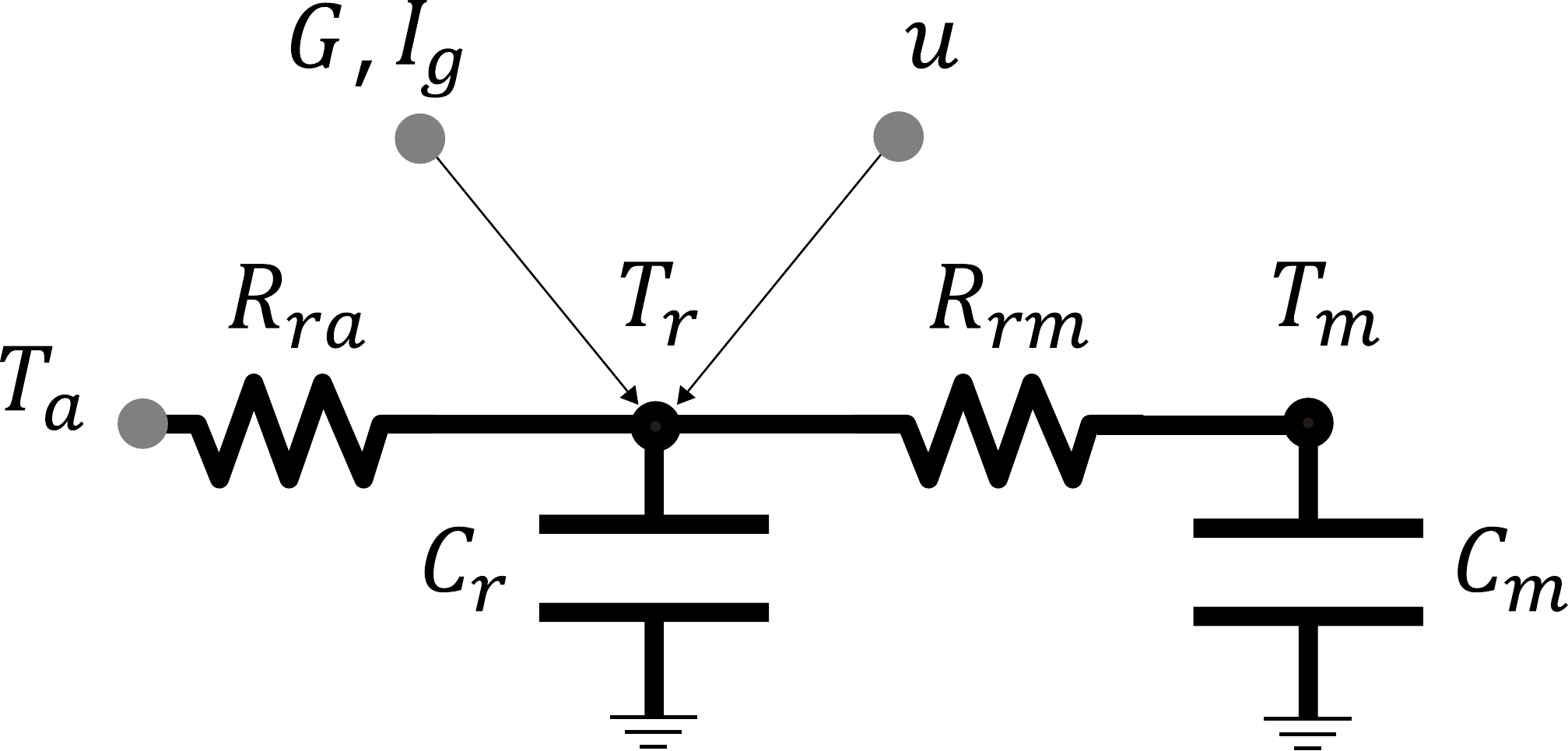}
    \caption{The resistance-capacitance network thermal model of the building}
    \label{fig:rcnet}
\end{figure}

\begin{equation}
\begin{split}
    \begin{bmatrix}
        \dot{T_r}\\
        \dot{T_m}
    \end{bmatrix} = \begin{bmatrix}
        -\left(\frac{1}{C_{r} R_{ra}} + \frac{1}{C_{r} R_{rm}}\right) & \frac{1}{C_{r} R_{rm}}\\
        \frac{1}{C_{m} R_{rm}} & -\frac{1}{C_{m} R_{rm}}
    \end{bmatrix} \cdot \begin{bmatrix}
        T_{r}\\
        T_{m}
    \end{bmatrix} \\ + \begin{bmatrix}
        b\\
        0
    \end{bmatrix} \cdot u + \begin{bmatrix}
        \frac{\alpha}{C_{r}} & \frac{\beta}{C_{r}} & \frac{1}{C_{r} R_{ra}} \\
        \frac{1-\alpha}{C_{m}} & \frac{1-\beta}{C_{m}} & 0 
    \end{bmatrix} \cdot \begin{bmatrix}
        G\\
        I_{g}\\
        T_{a}
    \end{bmatrix}.
\end{split}
\label{eq:2r2c}
\end{equation}
Here, $T_{r}$, $T_{m}$ and $T_{a}$ are the room temperature, temperature of building’s thermal mass and outside temperature respectively, $G$, $I_g$ represent solar irradiance and internal heat gains, and $R_{i}$, $C_{j}$ correspond to heat transfer parameters of the building. The room temperature $T_r$  is an observable state of the system that can be measured. Contrarily, $T_m$  is a hidden state of the system which cannot be measured directly and, in most cases, is extremely difficult to estimate. This modeling approach hence leads to a partially observable model of the building. Additionally, a low-level back-up controller is assumed which ensures that the room temperature remains within a predefined set of limits based on the comfort of the user. The action of this back-up controller affects the actual power consumption ($u^{\text{phys}}_{i}$) which is modeled as:
\begin{equation}
    u^{\text{phys}}_{i} =   \begin{cases}
                                0       &:      T_{r,i} > T^{\text{max}}_{r} \\
                                u_{i}   &:      T^{\text{min}}_{r} \leq T_{r,i} \leq T^{\text{max}}_{r}\\
                                u^{\text{max}} &: T_{r,i} < T^{\text{min}}_{r}\\
                            \end{cases}
\label{eq:backup}
\end{equation}
This backup controller ensures the comfort of the user and its actions leads to a difference between the power demanded ($u_{i}$) and the actual power consumed ($u^{\text{phys}}_{i}$). To solve \eqsref{eq:2r2c}{eq:backup}, an accurate estimate of hidden state~($T_{m}$) is required along with accurate measurements related to exogenous quantities like $G$ and $I_{g}$. Since in practice precise estimates, measurements are difficult to obtain, we will eventually get only an approximate solution. Further, the building parameters like conductivity of different walls change over time due to deterioration and lead to model bias. Hence modeling a household directly using \eqsref{eq:2r2c}{eq:backup} is a difficult and expensive process and can lead to biased and sub-optimal control policies. 

\subsection{Physics Informed Neural Network Configurations}
The state-space model defined in \equref{eq:2r2c} is a continuous time model. To map this model as an MDP, we discretize it based on the frequency of choosing an action ($u$). For our case study, we set this frequency to one action every 30 minutes and assume that the inside room temperature and power consumed by the heating source are monitored over this fixed time interval~($\Delta t$). The objective of the physics informed neural network model is to predict the room temperature and the power consumed for subsequent time steps. \\
Both architectures shown in \figref{fig:pinn-arch} were used and prior physics information was given by discretizing \equref{eq:2r2c}. This state-space model leads to a partially observable system. To mitigate this, input in the form of a sequence of past $k$ observable states and actions ($\textbf{x}^{\text{f}}_{i}$) along with observable state, actions and other exogenous information in the form of time of day~($t$) and outside air temperature is used. With these inputs, the networks predict the room temperature, power consumption and estimate the temperature of building thermal mass~($T_m$). The resulting state and action definitions are summarized in~\tabref{tab:notation-equ}.

\begin{table}[t]
\centering
\caption{State and Action Definitions for time step $i$}
\begin{tabular}{c c}
    \toprule
     Symbol & Physical Meaning\\
    \midrule
    $\textbf{x}^{f}_{i}$ & \{$(T_{r, i-k}, \ldots, T_{r, i-1})$, \\
                   & \quad $(u^{\text{phys}}_{i-k-1}, \ldots, u^{\text{phys}}_{i-2})$\}\\
    \midrule
    $\textbf{x}^{\text{obs}}_{i}$ & $(T_{r, i}, u^{\text{phys}}_{i-1})$ \\
    \midrule
    $\textbf{w}_{i}$ & $(t_i, T_{a,i})$\\
    \midrule
    $\textbf{z}_{i}$  & $T_{m, i}$\\
    \midrule
    $u_{i}$             &  Controller Power \\
    \bottomrule
\end{tabular}    
\label{tab:notation-equ}
\end{table}

The parameter $k$, referred to as `depth', controls the amount of information given to the neural network. It is important to note that the observed states for time step $i$ consist of the room temperature at this time step~($T_{r,i}$) and the actual power that was consumed during the last time step~($u^{\text{phys}}_{i-1}$). Prior physics knowledge is provided to these architectures by directly plugging in \equref{eq:2r2c} in the form of 
\begin{equation}
\begin{split}
    \begin{bmatrix}
        \dot{T_r}\\
        \dot{T_m}
    \end{bmatrix} &= \begin{bmatrix}
        -a_{11} & a_{12}\\
        a_{21} & -a_{22}
    \end{bmatrix} \cdot \begin{bmatrix}
        T_{r}\\
        T_{m}
    \end{bmatrix} \\ & \quad + \begin{bmatrix}
        b\\
        0
    \end{bmatrix} \cdot u + \begin{bmatrix}
        c_{11} & c_{12} & c_{13} \\
        c_{21} & c_{23} & 0 
    \end{bmatrix} \cdot \begin{bmatrix}
        0\\
        0\\
        T_{a}.
    \end{bmatrix}
\end{split}
\label{eq:pinn-phy-block}   
\end{equation}
The parameters $a_i$, $b$, $c_j$ are building specific parameters and initialized based on the building EPC values and further tuned during the training phase. This ensures that in the absence of accurate values of these parameters, the model can be initialized with approximate values. Moreover, these values can be tuned over time, thus taking into account any natural variations. It should be noted that this prior knowledge can be provided based on any other model of choice. However, as we focus on obtaining control-oriented models, a 2R2C modeling approach was chosen, which has been used previously in developing MPC-based control strategies~\cite{expt-mpc,Vrettos2018}. With this information setting, the different components of the loss function defined in~\equref{eq:gen-loss} can be formulated as:
\begin{equation}
\begin{split}
    \mathcal{L}_{\text{reg}} &= \frac{1}{N} \sum_{i=1}^{N} ({T}_{r,i} - \hat{T}_{r,i})^{2}\\
    & \quad + \frac{1}{N} \sum_{i=1}^{N} ({{u}^{\text{phys}}_{i}} - {\hat{u}^{\text{phys}}_{i}})^{2},\\
    \mathcal{L}_{\text{phys}} &= \frac{1}{N} \sum_{i=1}^{N} ({T}^{\mathcal{M}}_{m,i} - \hat{T}_{m, i})^{2}, \\
\end{split}
\label{eq:2r2c-pinn}
\end{equation}
The hidden state ($T^{\mathcal{M}}_{m,i}$) represents the physics module output and is computed by first estimating $\dot{T}_{r,i}$ and then using \equref{eq:pinn-phy-block} to obtain $T^{\mathcal{M}}_{m,i}$ as shown in \equref{eq:est_Tm}.
\begin{equation}
\begin{split}
    \dot{T}_{r,i} &= \frac{T_{r, i+1} - T_{r,i}}{\Delta t}\\
    T^{\mathcal{M}}_{m,i} &= \frac{1}{a_{12}} (\dot{T}_{r,i} + a_{11}\hat{T}_{r,i}  - b\hat{u}^{\text{phys}}_{i} - c_{13}T_{a, i})
\end{split}
\label{eq:est_Tm}
\end{equation}
Here, $\hat{T}_{r,i}$ is obtained as an output by using input sample $i-1$ and $\hat{u}^{\text{phys}}_{i}$ is obtained by using input sample $i$. The target value for the hidden state is explicitly dependent on the predictions of the room temperature and the power consumed. The loss functions defined in \equref{eq:2r2c-pinn} guides the outputs of the networks towards physically relevant values. 

\subsection{Training Data}
\label{sec:exp-setup:data}

We considered two different scenarios for obtaining training data for these architectures: (1)~a \emph{simulated} single household environment, and (2)~\emph{real-world cold storage} data. The simulated scenario has been specifically designed to assess the capacity of the proposed physics informed neural network architectures to estimate the hidden state $T_m$ of a building. After establishing this, later experiments focused on the real-world data scenario and assessed the performance of our proposed architectures for different configurations. Both scenarios involved observations related to room temperature, actual power consumption, control actions and outside air temperature. The frequency of these measurements was set to 1~measurement per 30~minutes. A training data-set equivalent to 120~days of such measurements was generated/collected. Similarly, a test data-set was generated equivalent to 5~days that were not a part of the training set. Each day corresponds to 48~input samples and the test sets for both scenarios used are shown in \figref{fig:data-scenarios}.

\subsubsection{Simulated Data}
\label{sec:exp-setup:data:simulated}
In this scenario, \eqsref{eq:2r2c}{eq:backup} were solved, as discussed in~\cite{Vrettos2018}. A discretization step of 1~minute was assumed and a control action was taken every 30~minutes. For every minute, a first order approximation of \equref{eq:2r2c} was solved to obtain the room temperature, hidden state and actual power consumption. To further simplify the scenario, we ignored the effects of solar irradiance and internal heat gain. The simulation was initialised by setting $T_r = T_m = 17 \text{\centigrades}$. Further, control action $u$ was chosen randomly and did not follow any active control logic. \Figref{subfig:simulated-data} shows a subset of data generated in this scenario.

\subsubsection{Real-world Data}
\label{sec:exp-setup:data:real}

\begin{figure}[t]
\centering
\subfigure[Simulated data for a household]{\includegraphics[width=2.35in]{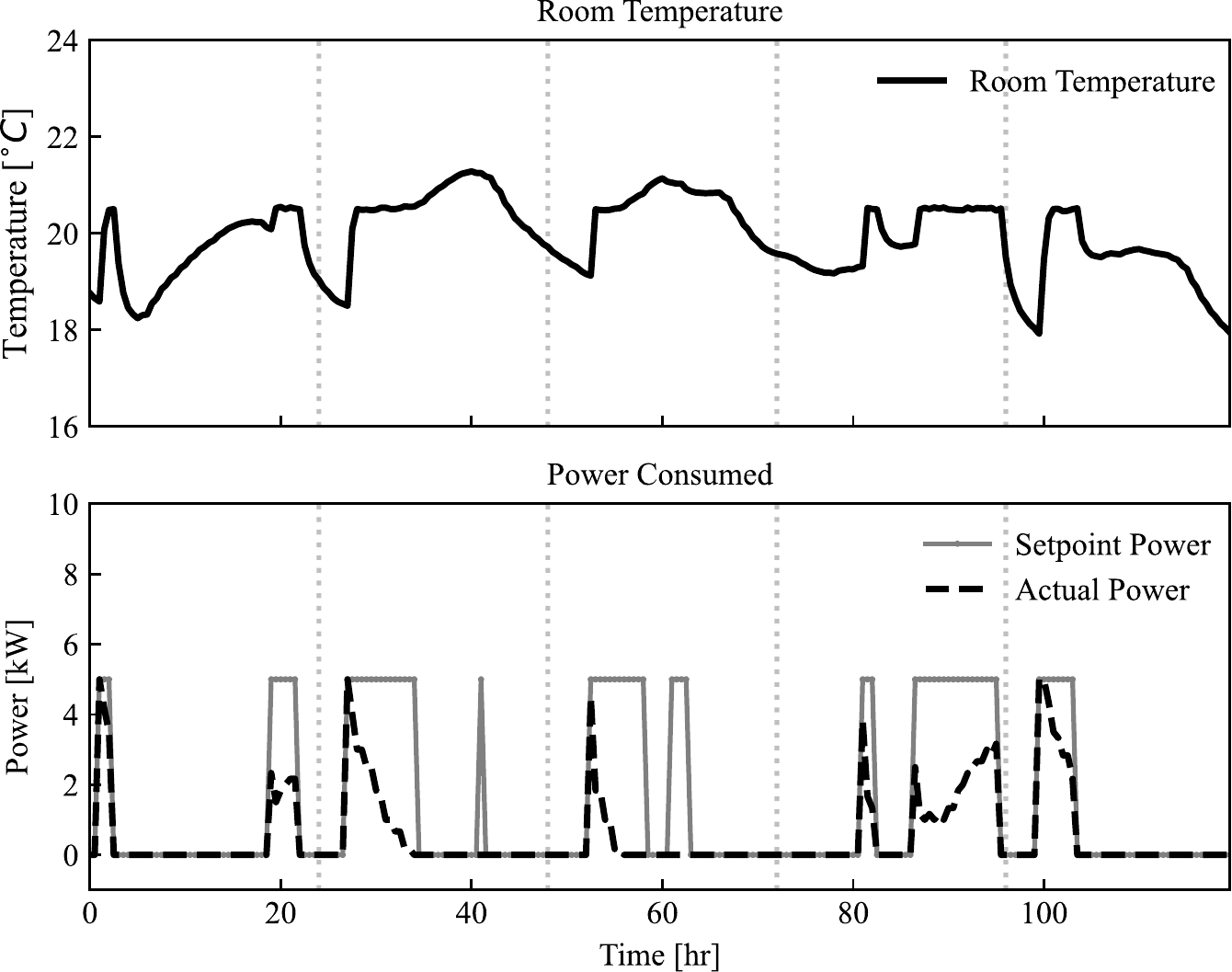}%
\label{subfig:simulated-data}}
\hfil
\subfigure[Real-world cold storage data]{\includegraphics[width=2.35in]{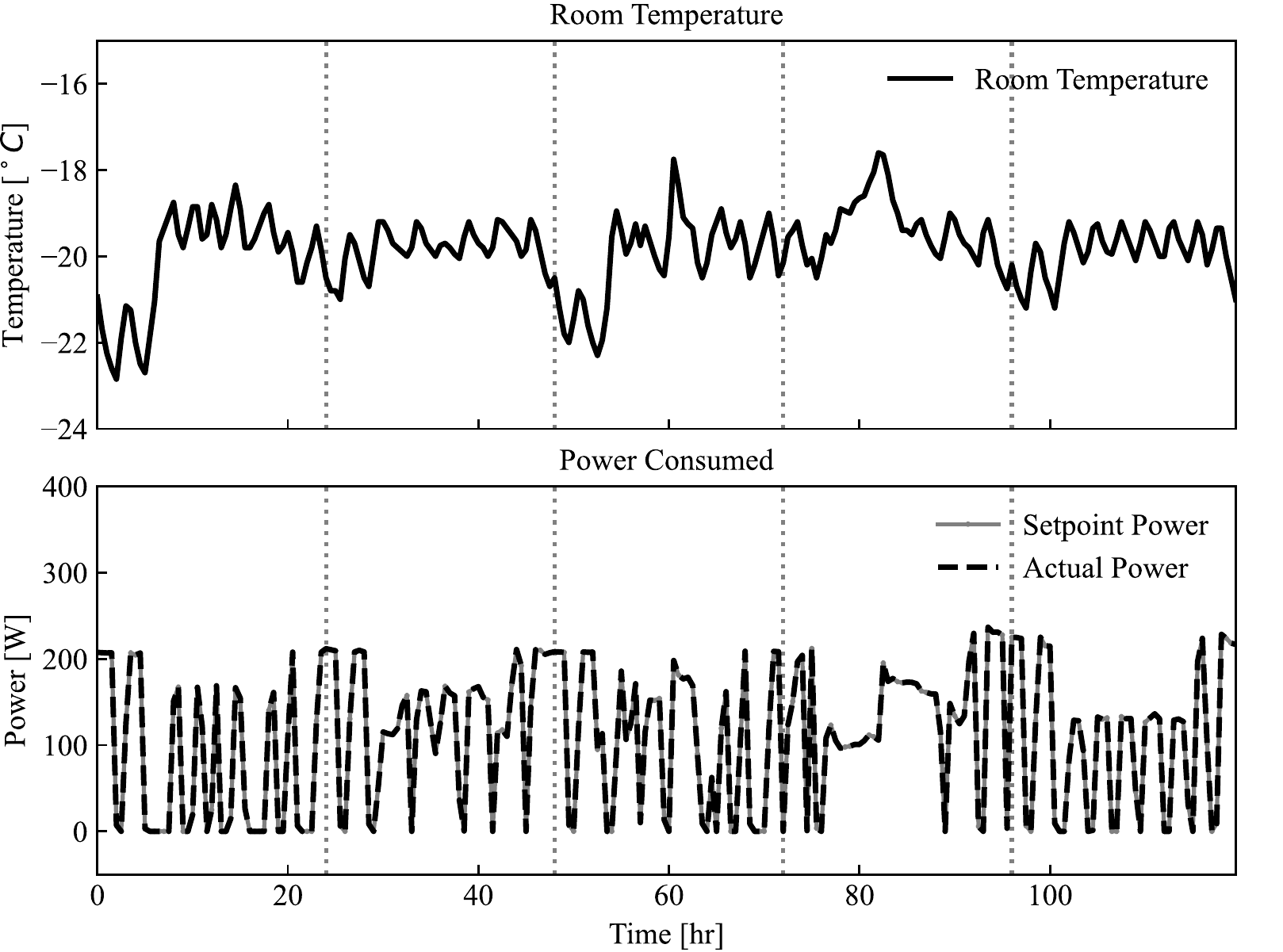}%
\label{subfig:cold-storage-data}}
\caption{Test data-sets used for both data scenarios, (a)~Simulated Data, (2)~Real-world scenario.}.
\label{fig:data-scenarios}
\end{figure}

This scenario involved data obtained from a cold storage. This scenario is more complex than the simulated data generated as it involved actions taken by an active control strategy and also included solar irradiance, internal heat gains, \etc The influence of these exogenous factors was recorded only indirectly, via the room temperature measurements, due to absence of sensors to directly measure them. \Figref{subfig:cold-storage-data} shows a subset of data corresponding to this scenario. In this cold storage case, no back-up controller was used and hence the actual power consumed is identical as the control setpoints.

\subsection{Parameter Tuning for Physics Informed Neural Network Architectures}

Besides different data scenarios, we also analyze the impact of different parameter configurations for both architectures. The input given to both architectures involves a sequence of past room temperatures and control actions. The length of this sequence, referred to as `depth’, determines the amount of past information available to the model and is an important parameter in the architecture. This information, to a certain level compensates for missing information like solar irradiance or internal heat gains, helping the model to better estimate the hidden state of the building~($T_m$). Further, the network sizes and hyperparameters~(learning rate, type of optimizers) were tuned for a base case of setting $\lambda=0$ for both architectures. This ensured that the network size and representative power was not constrained by the physics-based regularization, and we can observe supplementary gains in performance after tuning $\lambda$. The set of hyperparameters selected for both these architectures are listed in \ref{sec:appendex_hp}. The hyperparameter values were obtained by minimizing the mean absolute error in predicted temperature as a performance metric. For each of these configurations, we train 20~seeded models and the results are expressed using the mean (and standard deviation) of these 20~models. Because of the low training sample regime, training multiple models ensures that we obtain a distribution of performance values, thus mitigating the effects of possible outliers due to under-fitting. Additionally, \ref{sec:appendix:training_time} includes information regarding the training time required for each configuration and details the hardware setup used. 


\section{Results and Discussions}
\label{sec:results}
Three different experiments were performed to test our proposed PhysNet and PhysReg MLP architectures~(\figref{fig:pinn-arch}) and assess their performance as a control-oriented model.

\subsection{Architecture Validation}
\label{sec:results:validation}

\begin{figure}[t]
\centering
\subfigure[PhysNet Architecture]{\includegraphics[width=2.5in]{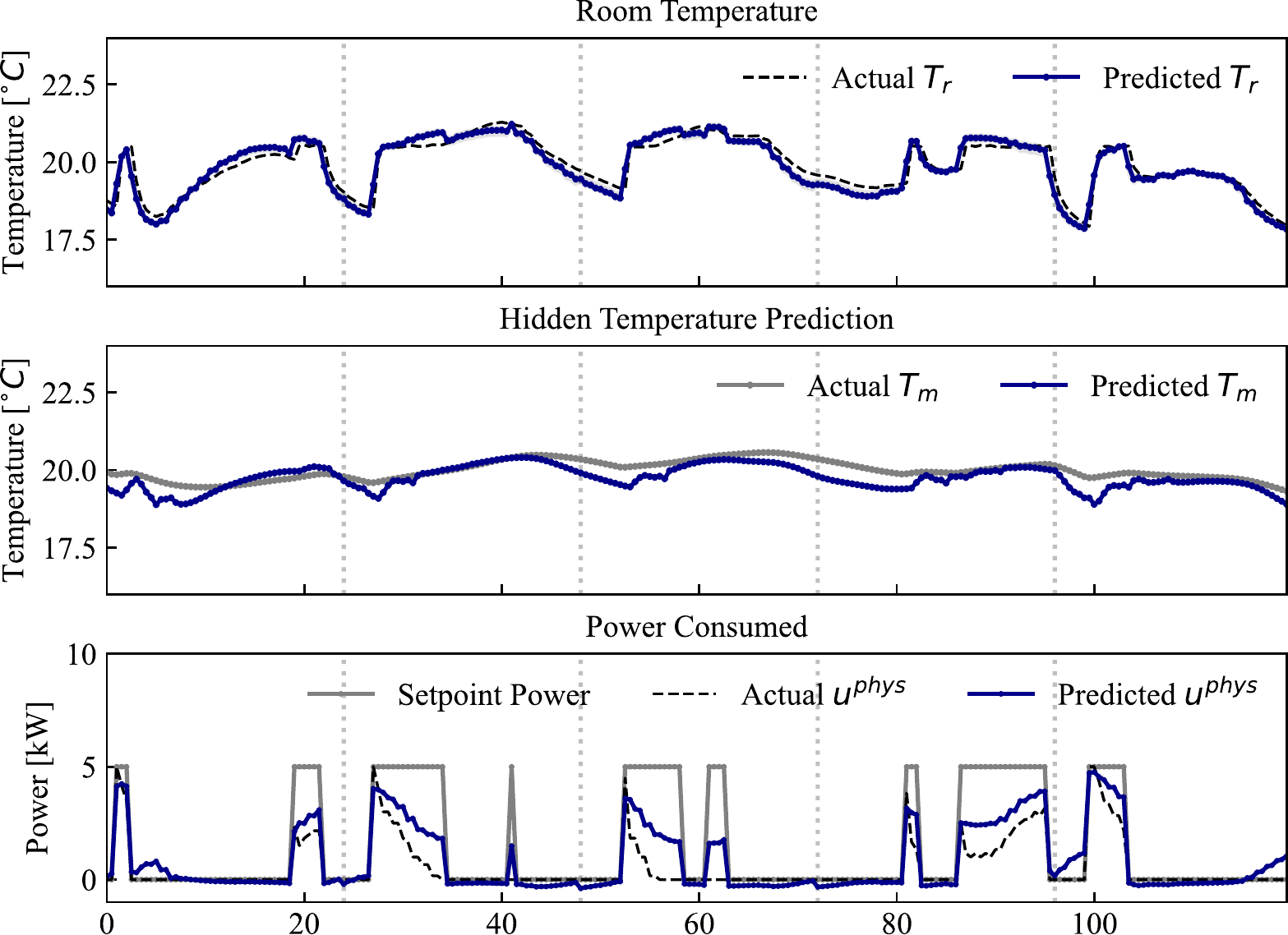}%
\label{subfig:res1-encoder-pinn}}
\hfil
\subfigure[PhysReg MLP Architecture]{\includegraphics[width=2.5in]{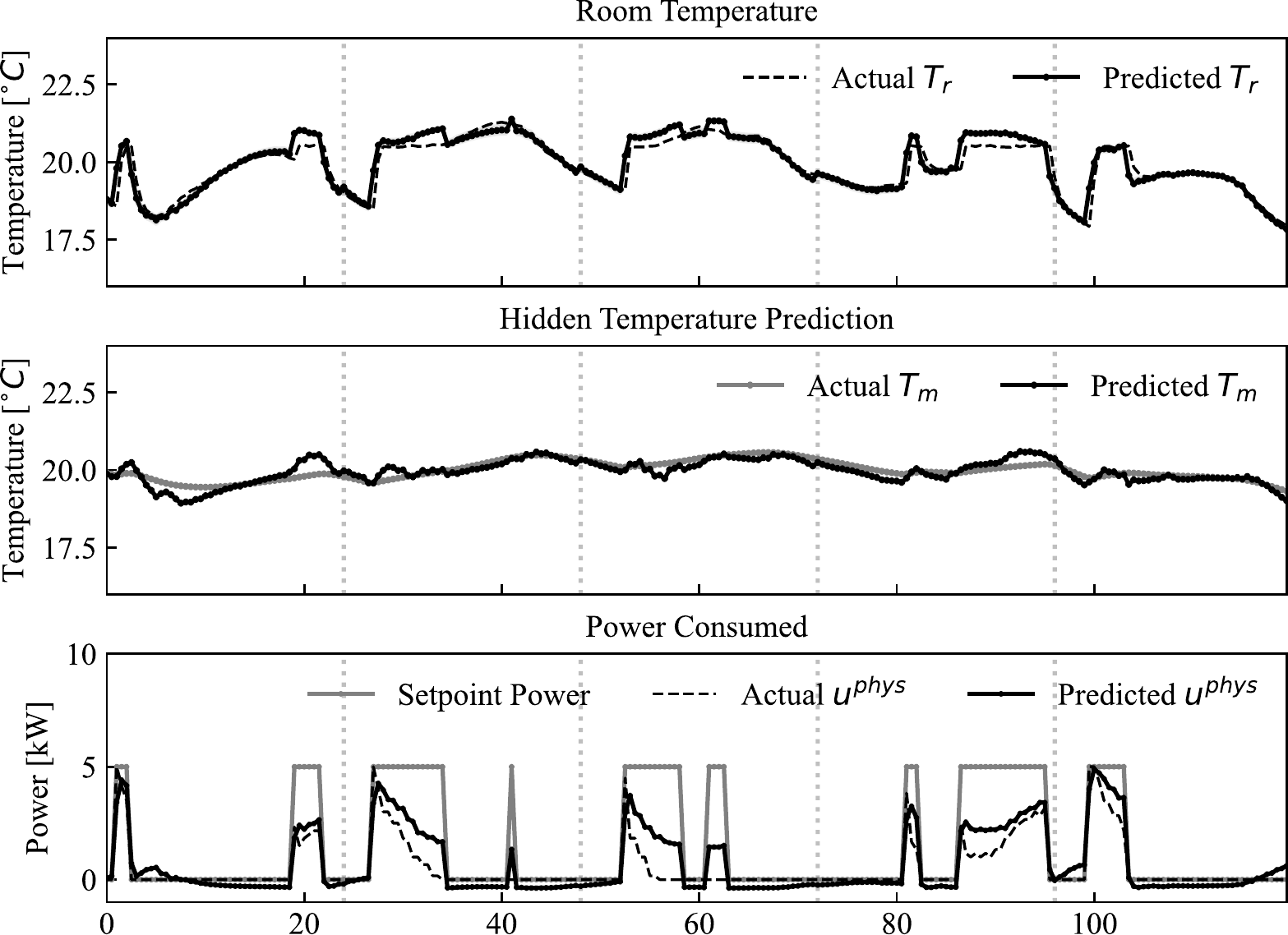}%
\label{subfig:res1-vanilla-pinn}}
\caption{Prediction results for the two physics informed neural network architectures on simulated data scenario}
\label{fig:res1-val}
\end{figure}

The aim of our first experiment was to validate the performance of the proposed physics informed neural network architectures in determining the quality of the hidden state estimates. For this purpose, simulated data was used for training and validation. The validation data-set, shown in \figref{subfig:simulated-data}, contains 240~samples for which we computed the Mean Absolute Error (MAE) of predicted room temperature~($T_r$) and predicted hidden state~($T_m$). A fixed training size of 120~days (5,760 samples) was used along with a fixed depth value of 8. \Figref{fig:res1-val} shows the predictions of both architectures.

We note that for both cases, the room temperature and action predictions follow the actual values closely, indicating a good prediction performance. Additionally, the estimates of $T_m$ track the actual values of hidden states, thus demonstrating the effectiveness of our proposed architectures for the given prediction task. \Tabref{tab:val_results} shows the MAE values for room temperature ($T_r$) and hidden state ($T_m$) predictions for this experiment.

\begin{table}[t]
    \centering
    \caption{Comparison of MAEs for room temperature~($T_r$) and hidden state~($T_m$) for proposed architectures}
    \begin{tabular}{c c c c }
    \toprule
        & MLP & PhysReg MLP & PhysNet \\
    \midrule
        $T_r$& $0.209^\circ{C}$ & $0.197^\circ{C}$ & $0.226^\circ{C}$\\
    \midrule
        $T_m$ & $1.413^\circ{C}$ & $0.385^\circ{C}$ & $0.436^\circ{C}$\\
    \bottomrule
    \end{tabular}
\label{tab:val_results}
\end{table}

\Tabref{tab:val_results} presents error values in \centigrades. A conventional MLP with the same hyperparameters as the PhysReg model was used to benchmark the performance of PhysNet and PhysReg MLP architectures. For all three networks, the mean errors are less than 0.25\centigrades, indicating a good performance. Comparing the architectures, the PhysReg model performs the best with an absolute error of 0.197\centigrades. However, there is a significant difference between mean errors for the hidden state, where conventional MLPs cannot estimate this state due to lack of target values, thus performing poorly in this metric. The physics informed neural network architectures perform $60-70 \%$ better than the conventional MLPs with an absolute error of less than 0.5\centigrades. These results demonstrate that PhysNet and PhysReg MLP architectures can be used effectively to predict room temperature and hidden state and hence are more suitable for control oriented thermal modeling of a building.

\subsubsection{Interpretability}
From the results presented in \tabref{tab:val_results}, it is evident that the proposed physics informed neural network architectures can effectively estimate the hidden state~($T_m$) while maintaining a good prediction accuracy for the observable state~($T_r$). It can be observed in \figref{fig:res1-val} that the estimate of temperature of building thermal mass acts as a thermal inertia quantity, having slower time dynamics compared to the room temperature. This behavior is consistent with our intuition and mimics the actual temperature of thermal mass. Thus, the trained physics-informed models give us additional insights about the behavior of the building by providing accurate estimates of both observable and hidden states of the building system. Such insights can help to better understand the predictions of the neural network and can be leveraged to design interpretable data-driven controllers.

\subsubsection{Real-world Data}
\begin{figure}[t]
\centering
\subfigure[PhysNet Architecture]{\includegraphics[width=2.5in]{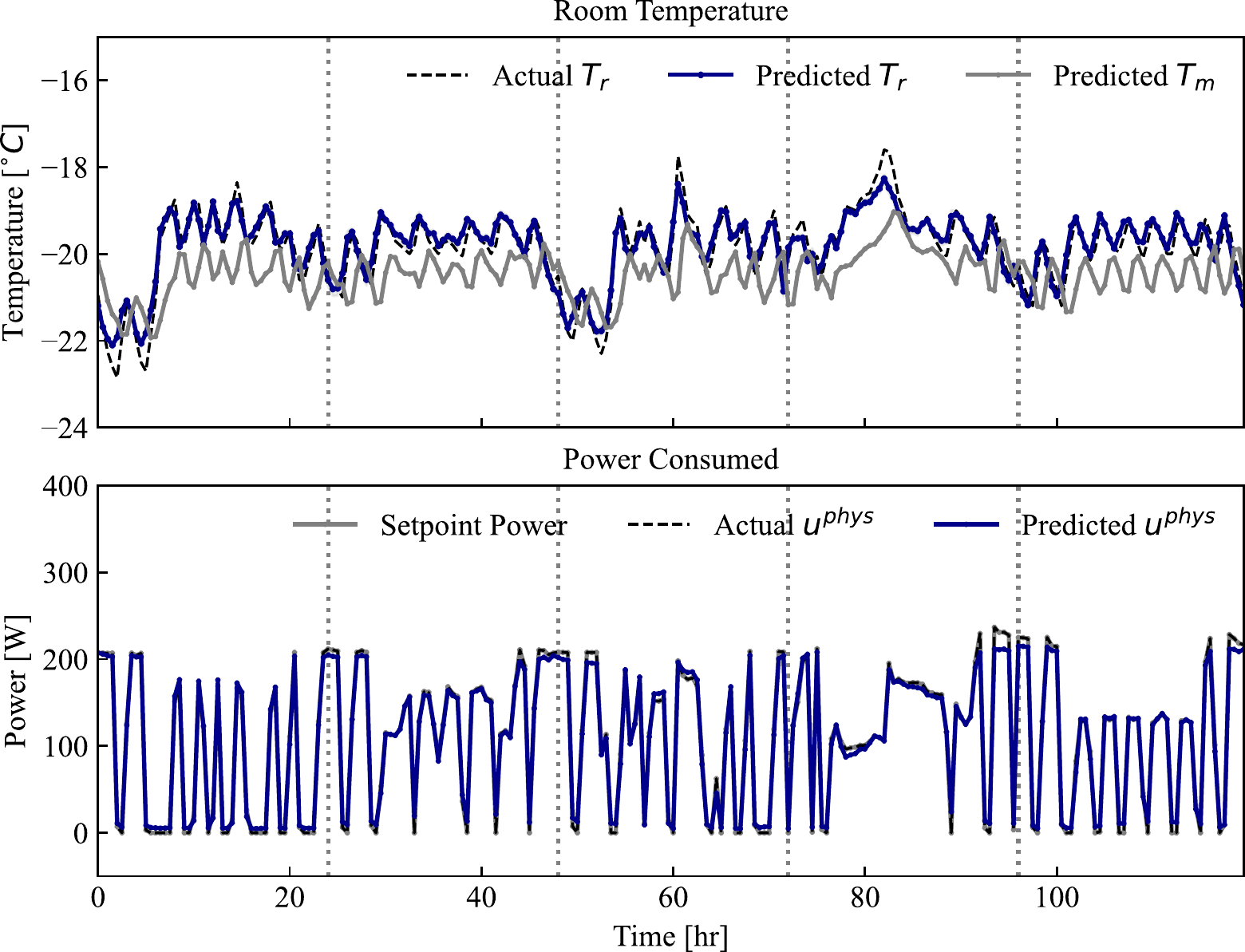}%
\label{subfig:res1-cs-physencoder}}
\hfil
\subfigure[PhysReg MLP Architecture]{\includegraphics[width=2.5in]{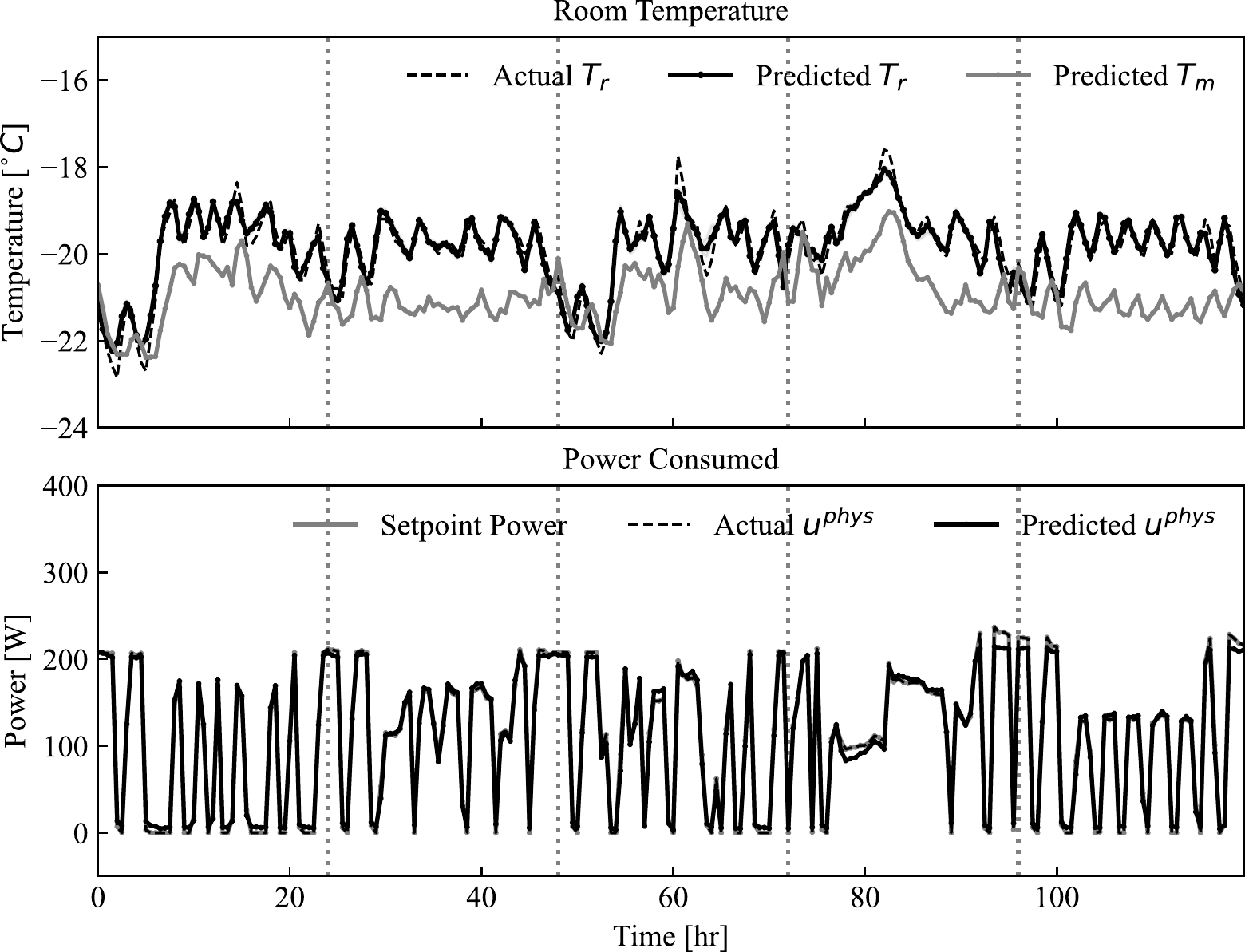}%
\label{subfig:res1-cs-physvanilla}}
\caption{Prediction results for physics informed neural network architectures on the real-world data scenario.}
\label{fig:res1-cs-val}
\end{figure}

Following these results, both physics informed neural network architectures were trained on real-world data obtained from a cold storage unit. Similar to the previous case, a training data size of 120 days was used and performance was validated on 5 test days, including a benchmark by a conventional neural network. \Figref{fig:res1-cs-val} shows the performance of PhysNet and PhysReg MLP architectures on the real-world data-set.

We note that both architectures accurately predict the room temperature and power consumption values for the 5~test days along with a plausible estimate of the hidden state of the system ($T_m$). The results shown in \figsref{fig:res1-val}{fig:res1-cs-val} and \tabref{tab:val_results} validate the performance of the proposed physics informed neural network architectures for the task of modeling the thermal behavior of a building. 

\subsection{Performance \vs Training Data Size}
\label{sec:results:training-data-size}

\begin{figure}[t]
\centering
\includegraphics[width=3.0in]{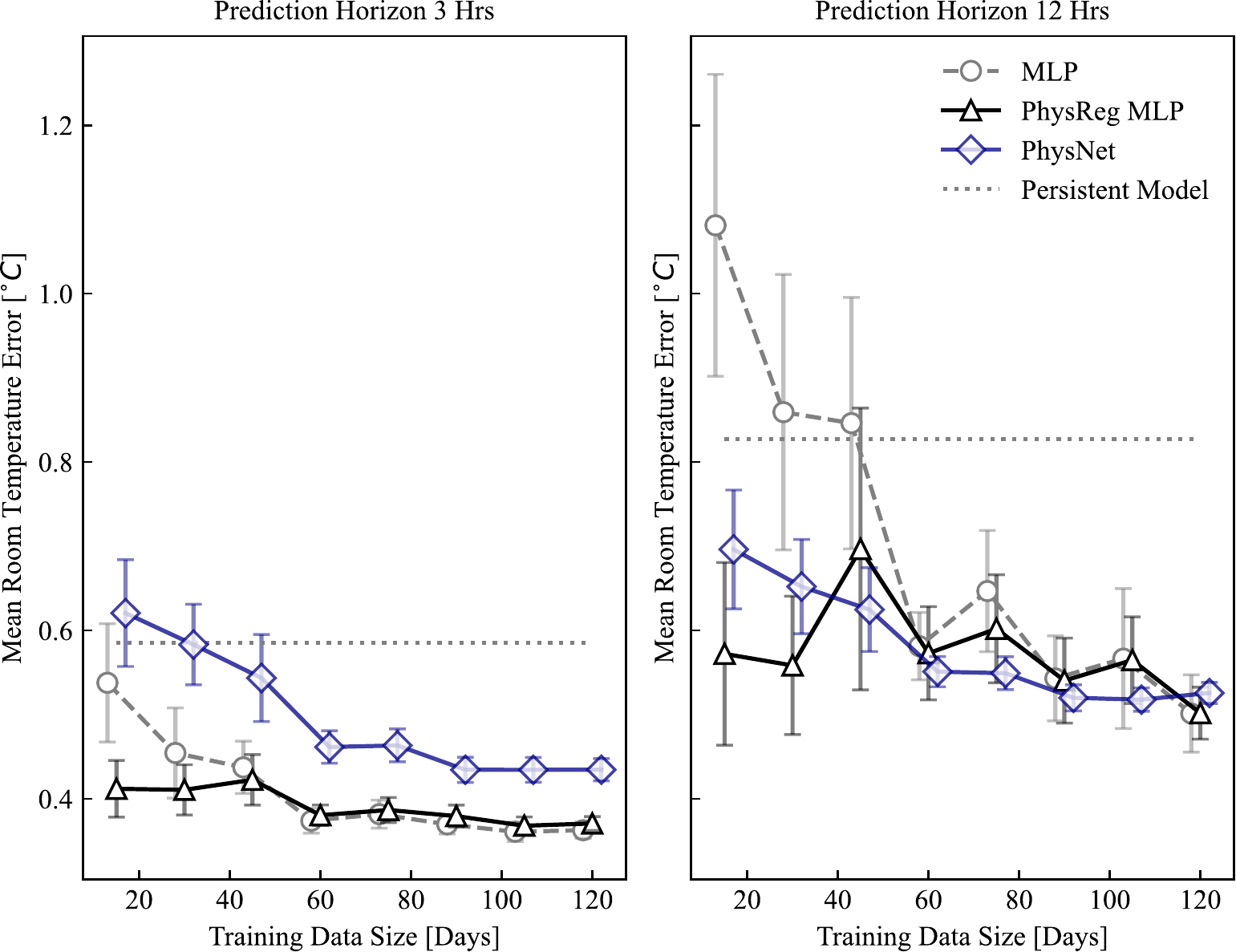}%
\caption{Mean room temperature prediction error for varying training data size. The plots represent mean error of 20~trained models and the error bars represent $\pm$ standard deviation. The models used in this experiment are trained and tested on real-world data obtained from a cold storage unit.}
\label{fig:res2}
\end{figure}

The second set of experiments analyzed the impact of training data size on the performance of physics informed neural network models. The motivation for using physics informed neural networks was to leverage prior knowledge to train models faster and more efficiently. To validate this, models were trained on real-world training data of varying size, sampled from the main training set. Each model was then tested using MAE as the performance metric on the test data-set of 240~samples (5~days) shown in \figref{subfig:cold-storage-data}. Two different test configurations were used, depending on the prediction horizon. Our architectures enable one-step ahead prediction. To obtain predictions for longer horizons, a recursive strategy was used, where the model output was fed back to the model as input to generate multi-step forecasts. This strategy mimics a tree search algorithm used in model-based RL techniques like~\cite{mu-zero2020}. The two test configurations used a prediction horizon of 3~hours (6~steps) and 12~hours (24~steps). The performance of physics informed neural networks was further compared to a conventional neural network and a persistence forecast model for both these configurations. \Figref{fig:res2} shows the model performance for different training data sizes for the real-world data scenario.

We note that for a prediction horizon of 12 hours, both PhysNet and PhysReg MLP architectures perform better than the conventional MLP. For smaller training data sizes (15-45~days) the predictions for physics informed neural network architectures attain an MAE that is at least 15\% lower than MLP. This difference decreases sharply with increasing training size, where for higher training sizes ($>$~90~days) the performance of all three architectures is similar. Contrary to this, for a shorter prediction horizon, the conventional MLP outperforms the PhysNet architecture, and performs similarly as the PhysReg MLP architecture. Additionally, in both configurations, all three architectures outperform a persistence forecasting model of similar prediction horizon for most training data sizes. This shows that introducing prior knowledge to the neural network architecture aids the network to learn more efficiently and requires less training data to reach equally good (or better) performance.

\subsection{Performance \vs Prediction Horizon Size}
\label{sec:results:horizon}

\begin{figure}[t]
\centering
\includegraphics[width=3.0in]{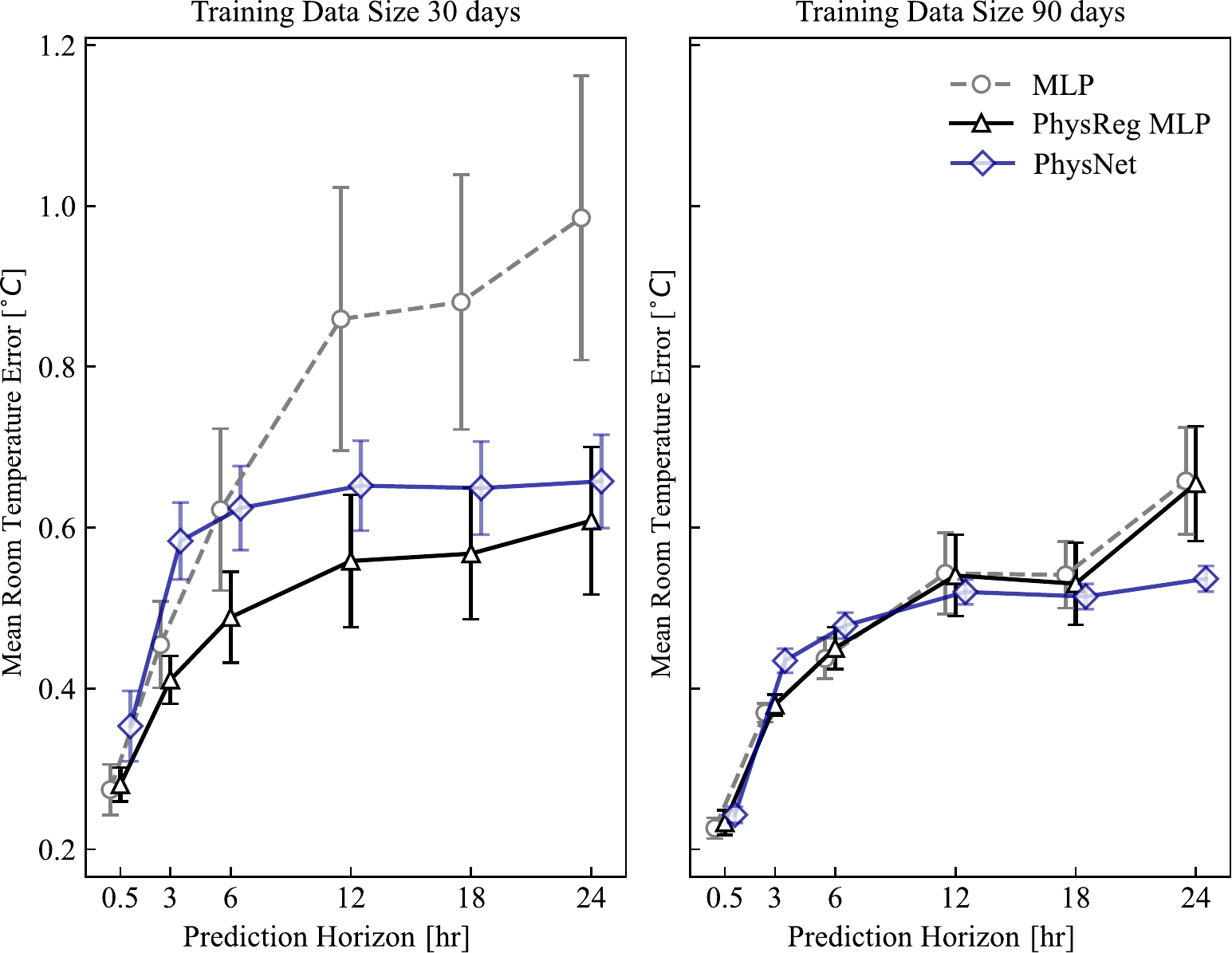}%
\caption{Mean room temperature prediction error for varying prediction horizons.The plots represent the mean error of 20~trained models and the error bars represent $\pm$ standard deviation. The models were trained and tested on the real-world cold storage data-set.}
\label{fig:res3}
\end{figure}

From \figref{fig:res2}, we note a difference in performance for different prediction horizons. While it is intuitively expected that increasing the prediction horizon will lead to compounding of errors, it is of interest to analyze how this performance degradation evolves for each of the two architectures. This experiment, thus analyzes the performance of physics informed neural networks for varying prediction horizons. Because of their relevance for typical control time frames, prediction horizons of \{0.5, 3, 6, 12, 18, 24\}~hours were selected, with each hour corresponding to 2~prediction steps. To include the impact of training data size, two training configurations of 30~days and 90~days were chosen. Like the previous experiment, real-world data was used with 5~test days as shown in \figref{subfig:cold-storage-data}. MAE of room temperature predictions was chosen as the performance metric and the performance was again benchmarked using a conventional MLP and a persistence forecast model with prediction horizon of 30~minutes (equalling 1~time step). \Figref{fig:res3} presents the results obtained for this experiment.

We note that for a large training data (120 days), all three architectures perform similarly in terms of mean values. However, the error bars indicate that PhysNet, PhysReg MLP architectures produce results with a more narrow distribution. This indicates a stable training performance in case of physics informed neural network architectures. For low training sample configurations, the performance of conventional MLP deteriorates rapidly with increase in prediction horizon size, with an error of close to 1\centigrades~for the case of 24~hours. While there is a significant decrease in performance for physics informed neural networks, the error margins remain around 0.75\centigrades~with a standard deviation of $\pm 0.3^\circ{\text{C}}$. This indicates that with less training data, the physics informed neural network models can use prior physics knowledge and lead to trained models that are stable and perform better than conventional MLPs. This is an important feature that can be leveraged in control applications for evaluating longer trajectories in tree searches.

These results demonstrate that introducing prior knowledge into a network leads to better predictions, makes the training process sample efficient and yields models that can be used for developing better control algorithms.

\section{Conclusion}
This work presented the application of physics informed neural networks for control-oriented thermal modeling of buildings. Our results show that both physics informed neural network architectures perform well for the given task of predicting the room temperature, with a low prediction error~(less than 0.25\centigrades). Further experiments confirm that physics informed neural networks are better suited for modeling in case of less training data and longer prediction horizons. This also indicates the robust training performance of PhysNets, PhysReg MLP architectures and their ability to generalize well even with fewer training samples. Additionally, we verify that the physics informed neural network models can estimate hidden states of the building effectively. This is an important feature and can be exploited further in developing control policies. Moreover, our proposed physics informed neural network architectures work with approximate values of model parameters and can incorporate partial knowledge about building physics. With such a setting, models for different buildings can be obtained using the same initial parameters and building physics, making this modeling approach scalable and easy to deploy.
\subsection*{Future Work}
Future work will involve two key directions:
\begin{enumerate*}[(i)]
    \item \label{it:control} Developing Control Algorithms, and 
    \item \label{it:arch} Improving PhysNet and PhysReg MLP architectures.
\end{enumerate*}
In~\ref{it:control} we will use these architectures in model-based RL algorithms like~\cite{mu-zero2020, dreamer}. The control agent will be capable of learning the model of the building and an optimum control policy simultaneously. Moreover, leveraging the learnt model, the agent can create a schedule for the next hours, making the decision making process interpretable and allowing human supervisory control. For~\ref{it:arch}, we aim to improve the architecture by introducing a direct multi-step forecasting capacity rather than the current one-step prediction setting. This will allow the architecture to produce one shot forecasts for a pre-defined prediction window. Other improvements include assessing the performance benefits of using recurrent neural networks in the model architecture and the role of clustering and transfer learning for scalable model deployment.

\section*{Funding}
This work was supported by the European Union's Horizon 2020 research and innovation programme under the projects BRIGHT (grant agreement no.\ 957816), RENergetic (grant agreement no.\ 957845) and BIGG (grant agreement no.\ 957047).

\appendix

\section{Hyperparameters for Physics Informed Neural Networks}
\label{sec:appendex_hp}
Both variants of physics informed neural networks were implemented using Pytorch Lightning package,~\cite{pl-light}. \Tabref{tab:hp-vanilla-pinn} and \tabref{tab:hp-encoder-pinn}  present the hyperparamters chosen for these architectures. $20$ models were training using the same set of hyperparameters and with different seeds between 1 to 20. Both architectures were trained using a batch size of 2048 and with 75 epochs. 

\begin{table}[h]
    \centering
    \caption{Hyperparameters for PhysReg MLP Architecture}
    \begin{tabular}{l c}
    \toprule
     Parameter & Value\\
    \midrule
        Optimizer & Adam\\
        Learning Rate & 0.001\\
        Activation Function & $\tanh$\\
        Batch Size  &  2048\\
        Hidden Layers & 2\\
        Neurons per layer & 64\\
    \bottomrule
    \end{tabular}
\label{tab:hp-vanilla-pinn}
\end{table}

\begin{table}[h]
\centering
\caption{Hyperparameters for PhysNet Architecture}
\begin{tabular}{l c}
    \toprule
     Parameter & Value\\
    \midrule
        Optimizer & Adam\\
        Learning Rate & 0.001\\
        Activation Function & $\tanh$\\
        Batch Size  &   2048 \\
    \midrule
    \multicolumn{2}{c}{\textit{Encoder Module} ($\theta_{L})$} \\
    \midrule
        Hidden Layers & 2\\
        Neurons per layer & 24\\
    \midrule
    \multicolumn{2}{c}{\textit{Dynamics Module} ($\theta_{d}$)} \\
    \midrule
        Hidden Layers & 1\\
        Neurons per layer & 128\\
    \bottomrule
\end{tabular}    
\label{tab:hp-encoder-pinn}
\end{table}

These hyperparameters were chosen using a grid search strategy and Mean Absolute Error for predictions on a validation set as the metric. The neural network hyperparameters were tuned first by setting $\lambda$ equal to 0. After this, the physics informed neural network parameters were tuned. \\ 
More information regarding the code can be found on: \url{https://github.com/GargyaGokhale/PhysNet_Thermal_Models}

\section{Training Time and Hardware Configuration}
\label{sec:appendix:training_time}
For results presented in \secref{sec:results:validation}, 20 instances of the neural network were trained using the same set of hyperparameters but different randomly initialised weight and bias values. On training, these 20 instances were used and the mean of their prediction was used. Table \ref{tab:training_time} presents the training time required for training these 20 instances for each of the two proposed physics informed neural network variants. \\
The training was carried out locally on a Dell laptop with Intel(R) Core(TM) i7-10850H CPU, 2.70GHz and 16 GB installed RAM.
\begin{table}[h]
\centering
\caption{Training Time Required}
\begin{tabular}{l c}
    \toprule
     Neural Network Type & Time\\
    \midrule
        PhysReg MLP &  3 minutes\\
        PhysNet & 4 minutes\\
    \bottomrule
\end{tabular}    
\label{tab:training_time}
\end{table}

\bibliographystyle{elsarticle-num} 
\bibliography{bibliograph-bib}

\begin{thebibliography}{10}
\expandafter\ifx\csname url\endcsname\relax
  \def\url#1{\texttt{#1}}\fi
\expandafter\ifx\csname urlprefix\endcsname\relax\def\urlprefix{URL }\fi
\expandafter\ifx\csname href\endcsname\relax
  \def\href#1#2{#2} \def\path#1{#1}\fi

\bibitem{ipcc-2021}
P.~Masson-Delmotte, V.and~Zhai, A.~Pirani, S.~Connors, C.~Péan, S.~Berger,
  N.~Caud, Y.~Chen, L.~Goldfarb, M.~Gomis, M.~Huang, K.~Leitzell, E.~Lonnoy,
  J.~Matthews, T.~Maycock, T.~Waterfield, O.~Yelekçi, R.~Yu, Z.~B.,
  \href{https://www.ipcc.ch/report/ar6/wg1/FullReport}{{IPCC, 2021: Climate
  Change 2021: The Physical Science Basis. Contribution of Working Group I to
  the Sixth Assessment Report of the Intergovernmental Panel on Climate
  Change}} (2021).
\newline\urlprefix\url{https://www.ipcc.ch/report/ar6/wg1/FullReport}

\bibitem{why-dr}
A.~Stawska, N.~Romero, M.~de~Weerdt, R.~Verzijlbergh, Demand response: For
  congestion management or for grid balancing?, Energy Policy 148 (2021)
  111920.

\bibitem{build-consump-sota2016}
X.~Cao, X.~Dai, J.~Liu,
  \href{http://dx.doi.org/10.1016/j.enbuild.2016.06.089}{{Building
  energy-consumption status worldwide and the state-of-the-art technologies for
  zero-energy buildings during the past decade}}, Energy and Buildings 128
  (2016) 198--213.
\newblock \href {https://doi.org/10.1016/j.enbuild.2016.06.089}
  {\path{doi:10.1016/j.enbuild.2016.06.089}}.
\newline\urlprefix\url{http://dx.doi.org/10.1016/j.enbuild.2016.06.089}

\bibitem{gen-review-2019}
T.~Q. P{\'{e}}an, J.~Salom, R.~Costa-Castell{\'{o}},
  \href{https://doi.org/10.1016/j.jprocont.2018.03.006}{{Review of control
  strategies for improving the energy flexibility provided by heat pump systems
  in buildings}}, Journal of Process Control 74 (2019) 35--49.
\newblock \href {https://doi.org/10.1016/j.jprocont.2018.03.006}
  {\path{doi:10.1016/j.jprocont.2018.03.006}}.
\newline\urlprefix\url{https://doi.org/10.1016/j.jprocont.2018.03.006}

\bibitem{mpc-basic}
J.~Drgo{\v{n}}a, J.~Arroyo, I.~C. Figueroa, D.~Blum, K.~Arendt, D.~Kim, E.~P.
  Oll{\'e}, J.~Oravec, M.~Wetter, D.~L. Vrabie, et~al., All you need to know
  about model predictive control for buildings, Annual Reviews in Control
  (2020).

\bibitem{mpc-challenges}
J.~C{\'\i}gler, D.~Gyalistras, J.~{\v{S}}iroky, V.~Tiet, L.~Ferkl, Beyond
  theory: the challenge of implementing model predictive control in buildings,
  in: Proceedings of 11th Rehva world congress, Clima, Vol. 250, 2013.

\bibitem{deep-rl-1}
S.~Brandi, M.~S. Piscitelli, M.~Martellacci, A.~Capozzoli, Deep reinforcement
  learning to optimise indoor temperature control and heating energy
  consumption in buildings, Energy and Buildings 224 (2020) 110225.

\bibitem{model-sota}
A.~Foucquier, S.~Robert, F.~Suard, L.~St{\'e}phan, A.~Jay, State of the art in
  building modelling and energy performances prediction: A review, Renewable
  and Sustainable Energy Reviews 23 (2013) 272--288.

\bibitem{EP-mpc}
G.~Gholamibozanjani, J.~Tarragona, A.~De~Gracia, C.~Fern{\'a}ndez, L.~F.
  Cabeza, M.~M. Farid, Model predictive control strategy applied to different
  types of building for space heating, Applied energy 231 (2018) 959--971.

\bibitem{modelica}
D.~Perera, D.~Winkler, N.-O. Skeie, Multi-floor building heating models in
  matlab and modelica environments, Applied Energy 171 (2016) 46--57.

\bibitem{mpc-real-problems}
E.~{\v{Z}}{\'a}{\v{c}}ekov{\'a}, Z.~V{\'a}{\v{n}}a, J.~Cigler, Towards the
  real-life implementation of mpc for an office building: Identification
  issues, Applied Energy 135 (2014) 53--62.

\bibitem{data-driven-narx}
F.~Ferracuti, A.~Fonti, L.~Ciabattoni, S.~Pizzuti, A.~Arteconi, L.~Helsen,
  G.~Comodi, Data-driven models for short-term thermal behaviour prediction in
  real buildings, Applied Energy 204 (2017) 1375--1387.

\bibitem{mpc-review-2021}
J.~Tarragona, A.~L. Pisello, C.~Fern{\'{a}}ndez, A.~de~Gracia, L.~F. Cabeza,
  {Systematic review on model predictive control strategies applied to active
  thermal energy storage systems}, Renewable and Sustainable Energy Reviews
  149~(May) (2021).
\newblock \href {https://doi.org/10.1016/j.rser.2021.111385}
  {\path{doi:10.1016/j.rser.2021.111385}}.

\bibitem{expt-mpc}
J.~{\v{S}}irok{\`y}, F.~Oldewurtel, J.~Cigler, S.~Pr{\'\i}vara, Experimental
  analysis of model predictive control for an energy efficient building heating
  system, Applied energy 88~(9) (2011) 3079--3087.

\bibitem{building-model2012}
I.~Hazyuk, C.~Ghiaus, D.~Penhouet, {Optimal temperature control of
  intermittently heated buildings using Model Predictive Control: Part I -
  Building modeling}, Building and Environment 51 (2012) 379--387.
\newblock \href {https://doi.org/10.1016/j.buildenv.2011.11.009}
  {\path{doi:10.1016/j.buildenv.2011.11.009}}.

\bibitem{swiss-mpc}
D.~Sturzenegger, D.~Gyalistras, M.~Morari, R.~S. Smith, Model predictive
  climate control of a swiss office building: Implementation, results, and
  cost--benefit analysis, IEEE Transactions on Control Systems Technology
  24~(1) (2015) 1--12.

\bibitem{dual_estimate}
S.~Baldi, S.~Yuan, P.~Endel, O.~Holub,
  \href{https://www.sciencedirect.com/science/article/pii/S0306261916301428}{Dual
  estimation: Constructing building energy models from data sampled at low
  rate}, Applied Energy 169 (2016) 81--92.
\newblock \href
  {https://doi.org/https://doi.org/10.1016/j.apenergy.2016.02.019}
  {\path{doi:https://doi.org/10.1016/j.apenergy.2016.02.019}}.
\newline\urlprefix\url{https://www.sciencedirect.com/science/article/pii/S0306261916301428}

\bibitem{mohak2020}
M.~Bhardwaj, S.~Choudhury, B.~Boots, Blending mpc \& value function
  approximation for efficient reinforcement learning, arXiv preprint
  arXiv:2012.05909 (2020).

\bibitem{data-driven-predictive2021}
A.~Kathirgamanathan, M.~{De Rosa}, E.~Mangina, D.~P. Finn,
  \href{https://doi.org/10.1016/j.rser.2020.110120}{{Data-driven predictive
  control for unlocking building energy flexibility: A review}}, Renewable and
  Sustainable Energy Reviews 135~(August 2020) (2021) 110120.
\newblock \href {http://arxiv.org/abs/2007.14866} {\path{arXiv:2007.14866}},
  \href {https://doi.org/10.1016/j.rser.2020.110120}
  {\path{doi:10.1016/j.rser.2020.110120}}.
\newline\urlprefix\url{https://doi.org/10.1016/j.rser.2020.110120}

\bibitem{alpha-go}
D.~Silver, A.~Huang, C.~J. Maddison, A.~Guez, L.~Sifre, G.~van~den Driessche,
  J.~Schrittwieser, I.~Antonoglou, V.~Panneershelvam, M.~Lanctot, S.~Dieleman,
  D.~Grewe, J.~Nham, N.~Kalchbrenner, I.~Sutskever, T.~Lillicrap, M.~Leach,
  K.~Kavukcuoglu, T.~Graepel, D.~Hassabis, Mastering the game of go with deep
  neural networks and tree search, Nature 529 (2016) 484--503.

\bibitem{rl-review}
J.~R. V{\'a}zquez-Canteli, Z.~Nagy, Reinforcement learning for demand response:
  A review of algorithms and modeling techniques, Applied energy 235 (2019)
  1072--1089.

\bibitem{sutton-barto}
R.~Sutton, A.~Barto, {An introduction to reinforcement learning}, Decision
  Theory Models for Applications in Artificial Intelligence: Concepts and
  Solutions (2011).
\newblock \href {https://doi.org/10.4018/978-1-60960-165-2.ch004}
  {\path{doi:10.4018/978-1-60960-165-2.ch004}}.

\bibitem{rl-fqi-ql}
L.~Yang, Z.~Nagy, P.~Goffin, A.~Schlueter, Reinforcement learning for optimal
  control of low exergy buildings, Applied Energy 156 (2015) 577--586.

\bibitem{mpc-vs-rl}
G.~Ceusters, R.~C. Rodr{\'\i}guez, A.~B. Garc{\'\i}a, R.~Franke, G.~Deconinck,
  L.~Helsen, A.~Now{\'e}, M.~Messagie, L.~R. Camargo, Model-predictive control
  and reinforcement learning in multi-energy system case studies, arXiv
  preprint arXiv:2104.09785 (2021).

\bibitem{rl-challenges}
Z.~Wang, T.~Hong, Reinforcement learning for building controls: The
  opportunities and challenges, Applied Energy 269 (2020) 115036.

\bibitem{trajectory-bert}
M.~Liu, S.~Peeters, D.~S. Callaway, B.~J. Claessens, Trajectory tracking with
  an aggregation of domestic hot water heaters: Combining model-based and
  model-free control in a commercial deployment, IEEE Transactions on Smart
  Grid 10~(5) (2019) 5686--5695.

\bibitem{control-models}
S.~Privara, J.~Cigler, Z.~V{\'a}{\v{n}}a, F.~Oldewurtel, C.~Sagerschnig,
  E.~{\v{Z}}{\'a}{\v{c}}ekov{\'a}, Building modeling as a crucial part for
  building predictive control, Energy and Buildings 56 (2013) 8--22.

\bibitem{control-models-2}
E.~Atam, L.~Helsen, Control-oriented thermal modeling of multizone buildings:
  Methods and issues: Intelligent control of a building system, IEEE Control
  Systems Magazine 36~(3) (2016) 86--111.
\newblock \href {https://doi.org/10.1109/MCS.2016.2535913}
  {\path{doi:10.1109/MCS.2016.2535913}}.

\bibitem{mu-zero2020}
J.~Schrittwieser, I.~Antonoglou, T.~Hubert, K.~Simonyan, L.~Sifre, S.~Schmitt,
  A.~Guez, E.~Lockhart, D.~Hassabis, T.~Graepel, T.~Lillicrap, D.~Silver,
  \href{http://dx.doi.org/10.1038/s41586-020-03051-4}{{Mastering Atari, Go,
  chess and shogi by planning with a learned model}}, Nature 588~(7839) (2020)
  604--609.
\newblock \href {http://arxiv.org/abs/1911.08265} {\path{arXiv:1911.08265}},
  \href {https://doi.org/10.1038/s41586-020-03051-4}
  {\path{doi:10.1038/s41586-020-03051-4}}.
\newline\urlprefix\url{http://dx.doi.org/10.1038/s41586-020-03051-4}

\bibitem{dreamer}
D.~Hafner, T.~Lillicrap, J.~Ba, M.~Norouzi, Dream to control: Learning
  behaviors by latent imagination, arXiv preprint arXiv:1912.01603 (2019).

\bibitem{pinn-og}
M.~Raissi, P.~Perdikaris, G.~E. Karniadakis,
  \href{https://doi.org/10.1016/j.jcp.2018.10.045}{{Physics-informed neural
  networks: A deep learning framework for solving forward and inverse problems
  involving nonlinear partial differential equations}}, Journal of
  Computational Physics 378 (2019) 686--707.
\newblock \href {https://doi.org/10.1016/j.jcp.2018.10.045}
  {\path{doi:10.1016/j.jcp.2018.10.045}}.
\newline\urlprefix\url{https://doi.org/10.1016/j.jcp.2018.10.045}

\bibitem{hamiltonian}
S.~Greydanus, M.~Dzamba, J.~Yosinski, Hamiltonian neural networks, in:
  H.~Wallach, H.~Larochelle, A.~Beygelzimer, F.~d\textquotesingle
  Alch\'{e}-Buc, E.~Fox, R.~Garnett (Eds.), Advances in Neural Information
  Processing Systems, Vol.~32, Curran Associates, Inc., 2019.

\bibitem{pinn-idlab-2021}
W.~Degroote, S.~{Van Hoecke}, G.~Crevecoeur, {Physics-Based Neural Network
  Models for Prediction of Cam-Follower Dynamics Beyond Nominal Operations},
  IEEE/ASME Transactions on Mechatronics PP~(c) (2021) 1.
\newblock \href {https://doi.org/10.1109/TMECH.2021.3101420}
  {\path{doi:10.1109/TMECH.2021.3101420}}.

\bibitem{pinn-heat-transfer2021}
S.~Cai, Z.~Wang, S.~Wang, P.~Perdikaris, G.~E. Karniadakis, {Physics-informed
  neural networks for heat transfer problems}, Journal of Heat Transfer 143~(6)
  (2021) 1--15.
\newblock \href {https://doi.org/10.1115/1.4050542}
  {\path{doi:10.1115/1.4050542}}.

\bibitem{physARMAX}
F.~Bünning, B.~Huber, A.~Schalbetter, A.~Aboudonia, M.~{Hudoba de Badyn},
  P.~Heer, R.~S. Smith, J.~Lygeros,
  \href{https://www.sciencedirect.com/science/article/pii/S0306261921017098}{Physics-informed
  linear regression is competitive with two machine learning methods in
  residential building mpc}, Applied Energy 310 (2022) 118491.
\newblock \href
  {https://doi.org/https://doi.org/10.1016/j.apenergy.2021.118491}
  {\path{doi:https://doi.org/10.1016/j.apenergy.2021.118491}}.
\newline\urlprefix\url{https://www.sciencedirect.com/science/article/pii/S0306261921017098}

\bibitem{phys_consistent}
L.~Di~Natale, B.~Svetozarevic, P.~Heer, C.~N. Jones, Physically consistent
  neural networks for building thermal modeling: theory and analysis, arXiv
  preprint arXiv:2112.03212 (2021).

\bibitem{pinn-multi-zone}
J.~Drgoňa, A.~R. Tuor, V.~Chandan, D.~L. Vrabie,
  \href{https://www.sciencedirect.com/science/article/pii/S0378778821002760}{Physics-constrained
  deep learning of multi-zone building thermal dynamics}, Energy and Buildings
  243 (2021) 110992.
\newblock \href {https://doi.org/https://doi.org/10.1016/j.enbuild.2021.110992}
  {\path{doi:https://doi.org/10.1016/j.enbuild.2021.110992}}.
\newline\urlprefix\url{https://www.sciencedirect.com/science/article/pii/S0378778821002760}

\bibitem{disentangle1}
Y.~Bengio, A.~Courville, P.~Vincent, {Representation learning: A review and new
  perspectives}, IEEE Transactions on Pattern Analysis and Machine Intelligence
  35~(8) (2013) 1798--1828.
\newblock \href {http://arxiv.org/abs/1206.5538} {\path{arXiv:1206.5538}},
  \href {https://doi.org/10.1109/TPAMI.2013.50}
  {\path{doi:10.1109/TPAMI.2013.50}}.

\bibitem{vae-disentangle}
C.~P. Burgess, I.~Higgins, A.~Pal, L.~Matthey, N.~Watters, G.~Desjardins,
  A.~Lerchner, Understanding disentangling in $\beta$-vae, arXiv preprint
  arXiv:1804.03599 (2018).

\bibitem{Vrettos2018}
E.~Vrettos, E.~C. Kara, J.~MacDonald, G.~Andersson, D.~S. Callaway,
  {Experimental Demonstration of Frequency Regulation by Commercial
  Buildings-Part I: Modeling and Hierarchical Control Design}, IEEE
  Transactions on Smart Grid 9~(4) (2018) 3213--3223.
\newblock \href {http://arxiv.org/abs/1605.05835} {\path{arXiv:1605.05835}},
  \href {https://doi.org/10.1109/TSG.2016.2628897}
  {\path{doi:10.1109/TSG.2016.2628897}}.

\bibitem{pl-light}
W.~Falcon, {The PyTorch Lightning team}, {PyTorch Lightning} (3 2019).
\newblock \href {https://doi.org/10.5281/zenodo.3828935}
  {\path{doi:10.5281/zenodo.3828935}}.

\end{thebibliography}

\end{document}